\documentclass[a4paper,fleqn,usenatbib]{mnras}

\usepackage[T1]{fontenc}
\usepackage{ae,aecompl}
\usepackage{float}

\usepackage{graphicx}	
\usepackage{amsmath}	
\usepackage{amssymb}	
\usepackage{caption}
\usepackage{subcaption}
\usepackage{multirow}
\title[KAT-7 observations of Sextans A  \& B]{H{\sc i} observations of Sextans A and B with the SKA pathfinder KAT-7}

\author[Namumba et al.]{
B.\ Namumba,$^{1}$\thanks{E-mail: brenda@ast.uct.ac.za}
C.\ Carignan, $^{1,2}$ and 
S.\ Passmoor $^{3}$
\\
$^1$Department of Astronomy, University of Cape Town, Private Bag X3, Rondebosch 7701, South Africa\\
$^2$Observatoire d$^{\prime}$Astrophysique de l$^{\prime}$Universit$\acute{e}$ de Ouagadougou, BP 7021, Ouagadougou 03, Burkina Faso \\
$^3$SKA South Africa, The Park, Park Road, Pinelands, 7705, South Africa\\
}
\date{Accepted 2018 April 19. Received 2018 April 10; in original form 2018 February 10}

\pubyear{2018}

\begin{document}
\label{firstpage}
\pagerange{\pageref{firstpage}--\pageref{lastpage}}
\maketitle

\begin{abstract}

H\textsc{i} observations of the Local Group dwarf irregular galaxies Sextans A and B, obtained with the Karoo Array Telescope (KAT-7) are presented. The KAT-7 wide field of view and excellent surface brightness sensitivity allows us to verify the true H\textsc{i} extent of the galaxies. We derive H\textsc{i} extends of 30$^{\prime}$ and 20$^{\prime}$ and total H\textsc{i} fluxes of 181 $\pm$ 2.0 Jy.km.s$^{-1}$ and 105 $\pm$ 1.4 Jy.km.s$^{-1}$ for Sextans A and B respectively. This result shows clearly the overestimate of the H\textsc{i} extent and total flux of 54$^{\prime}$ and 264 Jy.km.s$^{-1}$ reported for Sextans A using the Effelsberg observations. Tilted ring models allow us to derive the rotation curves (RCs) of Sextans A and B out to 550$^{\prime \prime}$ ($\sim$ 3.5 kpc) and 650$^{\prime \prime}$ ($\sim$ 4 kpc) respectively. The RCs of the two galaxies are seen to decline in the outer parts. The dark matter distribution in Sextans A is better described by the pseudo-isothermal halo model when a M/L ratio of 0.2 is used. For Sextans B, the mass model fits are not as good but again an isothermal sphere with a M/L of 0.2 represents best the data. Using the MOdified Newtonian Dynamics (MOND), better fits are obtained when the constant a$_{0}$ is allowed to vary. The critical densities for gravitational instabilities are calculated using the Toomre-$Q$ and cloud-growth based on shear criterion. We find that in regions of star formation, the cloud growth criterion based on shear explains better the star formation in both Sextans A and B.  
\end{abstract}




\section{Introduction}
The neutral hydrogen (H\textsc{i}) component of the interstellar medium (ISM) is an ideal tracer of the kinematics of disk galaxies. The ISM in nearby dwarf galaxies are the closest analog to conditions which prevailed in the early universe due to their low metallicity and high gas content. Numerous previous observations of dIrrs showed extremely large H\textsc{i} envelopes several times larger than their optical cores (e.g.,\ \citealp{1981A&A...102..134H,1996AJ....111.1551M,1998AJ....116.2363W,2007MNRAS.375..199G,2011AJ....141..204K,2011AJ....142..173H,2014A&A...561A..28S,2017arXiv170809447N}). Such extended H\textsc{i} envelopes allow us to probe very far out into the dark halo potential and derive the large-scale kinematics of these galaxies (e.g.,\ \citealp{1998ApJ...506..125C}). 

Most interferometric studies of H\textsc{i} gas in dwarf galaxies lack the required sensitivity to detect large scale extended H\textsc{i} emission (e.g. \citet{1986A&A...165...45S,2000AJ....120.3027C,2000ApJ...537L..95D,2011AJ....141..204K,2012AJ....144..134H}). We would ideally probe the extended H\textsc{i} gas using single dish telescopes as these instruments do not filter out any emission \citep{2014AJ....147...48P}. However, single dish telescopes do not have the required spatial resolution to derive the kinematics of the gas. Our best alternative is to use an interferometer with short baselines that has better spatial resolution and is sensitive to large scale structures.

As we prepare for the Square Kilometer Array (SKA) in the near future, SKA pathfinders are being built for engineering tests and early science. Among the telescopes being built for the SKA early science is MeerKAT. One of the main aim of MeerKAT is to detect low column density gas in nearby galaxies using high resolution and achieving column density sensitivity of $\leqslant 10^{18}$ cm$^{-2}$. As MeerKAT is still being built, the KAT-7, initially built as an engineering test-bed for MeerKAT has been able to produce new exciting early SKA science. The KAT-7 compact configuration of its seven 12m dishes, combined with the low $T_{sys} \sim$ 26 K have been able to detect, when present, low column density gas in many nearby galaxies that could not be detected by other synthesis arrays such as the VLA or ATCA. 

KAT-7 observations of the Magellanic-type spiral NGC 3109 \citep{2013AJ....146...48C} detected 40$\%$ more H\textsc{i} mass than what was detected with the VLA observations. This allowed the measurement of the RC of NGC 3109 out to 32$^{\prime}$, doubling the angular extent of existing measurements. \citet{2015MNRAS.450.3935L} detected 33$\%$ more flux for NGC 253 than what was previously detected with the VLA, giving new results on the kinematics of the low column density gas in NGC 253. KAT-7 polarized radio continuum and spectral line observations of M83 revealed massive H\textsc{i} gas distribution that appears to be tightly coupled with interaction of the galaxy and the enviroment \citep{2016MNRAS.462.1238H}. These observations detected more flux than the VLA observations and allowed the rotation curve to be measured out to 50 kpc. The recently published KAT-7 observations of the nearby dwarf irregular galaxy NGC 6822 \citep{2017arXiv170809447N} revealed 23$\%$ more H\textsc{i} mass than the previous ATCA observations. 

The discrepancy between the single dish (Effelsberg and Green Bank Telescope (GBT)) data on the true H\textsc{i} extent of Sextans A \citep{1981A&A...102..134H,2011AJ....142..173H} and the interferometric data from the Very Large Array (VLA) \citep{1988A&A...198...33S,2002AJ....123.1476W} has motivated us to map Sextans A and B in H\textsc{i} using the KAT-7. KAT-7 is ideally suited for this project due to its short baselines and low receiver temperature which makes it sensitive to large scale, low surface brightness emission. 

The structure of the paper is as follows. Brief descriptions of Sextans A and B are given in Section 2, followed by the observations and data reduction in Section 3. In Section 4, we present the results on the H\textsc{i} distribution. The rotation curve results derived from the tilted ring fit are presented in Section 5, and the dark matter mass models of the galaxies are explained in Section 6. In section 7, we explore instability models for the onset of star formation in Sextans A and B and in Section 8 we summarize our work.

\section{Sextans A and Sextan B}
Sextans A (DDO 75, UGC 205) is a dIrr galaxy at a distance of 1.3 Mpc \citep{1991rc3..book.....D}, which places it in the Local Group of galaxies, lying between Sextans B and NGC 3109. Sextans A is classified as an IBm galaxy \citep{1991rc3..book.....D}, having an absolute magnitude of M$_{V} \sim$ -14 \citep{2005AJ....130.1558K}. The stellar distribution in Sextans A has been observed to have a square-like distribution \citep{1996ApJ...462..732H}. Its optical angular diameter is D$_{0}$ = 4$^{\prime}$.8 \citep{1991rc3..book.....D}. 

\citet{1981A&A...102..134H} first mapped the H\textsc{i} distribution around Sextans A using the  Effelsberg 100m single dish telescope. Their results yielded a large H\textsc{i} extent of 54$^{\prime}$, which is 5.8 times the galaxy optically-defined Holmberg diameter, and a total H\textsc{i} mass of 1.6$\times 10^{8}$ M$_{\odot}$. The first interferometric observations of Sextans A were done by \citet{1988A&A...198...33S} using the VLA. They were able to detect only about 50$\%$ of the total H\textsc{i} flux reported by \citet{1981A&A...102..134H} and measured an H\textsc{i} extent of 9.4$^{\prime}$. \citet{1988A&A...198...33S} suggested that the 50$\%$ missing gas could exists in large scale low column density gas unable to be detected by the VLA due to the lack of short baselines and limited field of view. \citet{2002AJ....123.1476W} mapped Sextans A using a VLA mosaic that allowed them to sample a much larger field of view than \citet{1988A&A...198...33S}. They were able to recover only 62$\%$ of the flux reported by \citet{1981A&A...102..134H} with an H\textsc{i} extent of 18$^{\prime}$. Comparing their VLA H\textsc{i} maps from the peaks in the profiles at each position with data from the H\textsc{i} Parkes All-Sky Survey (HIPASS) observations confirmed the fact that there was a good deal of H\textsc{i} gas missing from the VLA mosaic due to the short spacings problem.

To verify the existence of the gas at the outer edges of Sextans A and explore how far it extends, \citet{2011AJ....142..173H} observed Sextans A with the Green Bank 100m single dish telescope (GBT). Their observations detected 25$\%$ more flux compared to \citet{2002AJ....123.1476W}, but only 78$\%$ of the flux detected by \citep{1981A&A...102..134H}. They measured an H\textsc{i} extent of 22.1$^{\prime}$. They outlined significant differences between the GBT and the Effelsberg observations that may contribute to the difference in the observed H\textsc{i} extent. One difference was that the signal to noise of the GBT observations was higher, and the extended features in the \citet{1981A&A...102..134H} occured at low signal to noise. The other difference was that the GBT has a clear point spread function out to several degrees from the main beam while the Effelsberg telescope has significant sidelobes. This lead them to believe that the Effelsberg map had galaxy emission entering from the sidelobes, which resulted in an overestimate of both the extent and the total flux.

Sextans B (DD0 70, UGC 5373) is classified as an IBm galaxy at a distance of 1.3 Mpc \citep{1991rc3..book.....D}. It is a faint galaxy with an absolute magnitude of M$_{V} \sim$ -14 \citep{2005AJ....130.1558K}. As far as global properties are concerned, Sextans B is considered to be a twin of Sextans A, except for the profile width, due to different inclinations. \citet{1981A&A...102..134H} first mapped the H\textsc{i} of Sextans B using the Effelsberg single dish telescope. They were able to derive an H\textsc{i} extent of 13$^{\prime}$, which is 1.7 times the Holmberg diameter, and a total H\textsc{i} mass of 1.3$\times10^{8}$M$_{\odot}$. VLA study of the neutral hydrogen in Sextans B \citep{2009AAS...21342432W} showed that the H\textsc{i} kinematics of this galaxy are similar to those of other dwarfs. They found that the rotation curve of Sextans B was flat out to a radius of 250$^{\prime \prime}$, which is about 1.6 kpc. Several distinct H\textsc{i} holes have also been identified in Sextans B. \citet{2015AJ....149..180O} derived the rotation curve of Sextans B out to a radius of $\sim$ 3 kpc using the VLA Local Irregulars That Trace Luminosity Extremes, The H\textsc{i} Nearby Galaxy Survey (LITTLE THINGS) \citep{2012AJ....144..134H}. 

\begin{table}
\scriptsize

\caption{\small Basic properties of Sextans A and B.}
\begin{minipage}{\textwidth}
\begin{tabular}{l@{\hspace{0.70cm}}c@{\hspace{0.05cm}}c@{\hspace{0.05cm}}}   
\hline

Parameter &Sextans A &  Sextans B \\
   
\hline \hline  
Morphology  & IBm$^{a}$&IBm$^{a}$ \\
Right ascension (J2000) &10:11:01.3$^{a}$& 09:59:59.9$^{a}$ \\
Declination (J2000)&-04:42:48.0$^{a}$& 05:19:57$^{a}$ \\
Distance (Mpc) & 1.3$^{a}$ &1.3$^{a}$\\
M$_{B}$ (Mag)& -13.9$^{d}$ & -13.9$^{d}$\\
V$_{\text{heliocentric}}$ (km.s$^{-1}$) & 324$^{a}$& 301$^{a}$\\
PA$\_$opt($\circ$) &41.0$^{e}$ & 88.0$^{e}$\\
Inclination$\_$opt($\circ$) & 33.5$^{e}$&57.8$^{e}$\\
Total HI mass (M$_{\odot}$)&(7.3 $\pm$ 0.07) $\times10^{7}$& (4.2 $\pm$ 0.06) $\times10^{7}$\\
\hline     \\
\multicolumn{3}{@{} p{8 cm} @{}}{\footnotesize{\textbf{Notes.} Ref\,(a) \citet{1991rc3..book.....D}; (b) \citet{2003AJ....126..187D}; (c) \citet{2002A&A...389..812K}; (d) \citet{2005AJ....130.1558K}; (e) \citet{2012AJ....144..134H}}}
\label{coords_table}
 
\end{tabular}   

\end{minipage}
\end{table}  
\section{KAT-7 Observations and Data Reduction}
The observations were obtained with the compact, seven-dish KAT-7 telescope \citep{2016MNRAS.460.1664F}, located close to the South African SKA core site in the Northern Cape Karoo desert region. The parameters of the KAT-7 observations are given in Table 2. 

The data were collected between 2014 December and 2015 July. Sextans A was observed over 19 observing sessions for a total of 60 hours on source while Sextans B was observed for 17 sessions for a total of 51 hours on source. The H\textsc{i} observations were carried out using the c16n25M4K spectral line mode. This correlator mode gives a total bandwidth of 12.5 MHz and 4096 channels of 0.64 km.s$^{-1}$ width. The central frequencies of our observations were 1418.9 MHz for Sextans A and 1419.0 MHz for Sextans B. All the data were hanning smoothed to 1.28 km.s$^{-1}$ before calibration. The observations were done using a 5, 6 or 7 antenna configuration depending on the availability of the antennas.

All the data were reduced using the standard calibration tasks in the Common Astronomy Software Application CASA 3.4.0 package \citep{2007ASPC..376..127M}. Calibration was performed separately for each observing session. Flagging of bad data due to radio frequency interferences (RFI) and antenna malfunctions were done using the CASA task \textsc{flagdata}. The corrections for the flux/bandpass shape were determined using the calibrator 0407+658. The time-varying phases and antenna gains were calibrated based on observations of the phase calibrator 0942-080. The calibration solutions were applied to the targets and the target sources were then split from the calibration sources using the task \textsc{split}. The individual calibrated data sets were then combined together using the CASA task \textsc{concat}. The combined data were continuum subtracted by using the first order polynomial in the task \textsc{uvcontsub}. Doppler corrections were performed on the data to ensure that proper velocity coordinates were used \citep{2013AJ....146...48C}. At this point, visibilities satisfying the condition $|u| \leqslant  10\lambda$ were removed from the data set to avoid the low level RFI that are correlated when the fringe rate is equal to zero \citep{2015MNRAS.450.3935L,2017MNRAS.464..957H}.

The data were Fourier transformed using the CASA task \textsc{clean}. Firstly, dirty cubes were imaged and the rms noise in line free channels was measured. For both Sextans A and B, the noise level in line free channels was found to be $\sim$ 1.5 times the theoretical noise, see Table 2. Using a flux threshold of 1 times the typical r.m.s of the flux in a line free channel, the data were cleaned in a non-interactive mode using a clearly defined mask. For each galaxy, two cubes were produced by applying the natural (na) and robust 0 weighting to the uv data. Taking into account the spatial resolution of KAT-7, we have decided to use the robust 0 cubes for our analysis. 

\subsection{Deriving H{\sc i} Maps}

For each galaxy, the data cube was smoothed to 2 times the original spatial resolution. Noise pixels were removed from the smoothed cubes by excluding pixels $\leqslant 2\sigma$, where $\sigma$ is the noise in line free channels. The remaining noise pixels were removed by creating masks around the galaxy emission in each channel. The final mask from the smoothed cube was then applied to the full resolutions cubes. The integrated H\textsc{i} maps were created by adding together all clipped channels with emission using a primary beam corrected cubes. The maps were converted to column density by using the following formula:
\begin{equation}
H\textsc{i}\ ({\rm cm^{-2}}) = \frac{1.835 \times 10^{18} \times dv \times 6.07 \times 10^{5}}{\theta_{major} \times \theta_{minor}}
\end{equation}
where $dv$ is the velocity resolution, $\theta_{major}$ is the major axis  and $\theta_{minor}$ is the minor axis of the beam. The values of these parameters are given in Table 2. 

To construct the velocity field maps, two methods were considered. The most popular one being the intensity weighted mean (IWM). However, this method is known to produce uncertainties in deriving velocities when the S/N is low \citep{2008AJ....136.2648D} as in our case for Sextans A. An alternative, in this case, was to construct the velocity field maps by fitting Gaussians to each profile. This method has already been implemented successfully by various authors \citep [see e.g.][]{1990AJ.....99..178C} and has proven to give a better result compared to the IWM. 

We fitted a first order Gaussian to each H\textsc{i} cube line profile to generate the velocity field maps. This was done using the GIPSY task \textsc{xgaufit}. We used the data sets without primary beam correction in order to retain the original properties of the data when performing profile fitting. To ensure high quality velocity field maps, three filters were used: 1) only profiles with fitted fluxes maxima higher than 3$\sigma$ were retained, 2) profiles with a fitted line width less than the velocity resolution of the data were excluded, and 3) fitted profiles had to be within the velocity range of the data. The dispersion maps were produced using the AIPS task \textsc{momnt}. The final velocity and dispersion maps were masked using the amplitude map from \textsc{xgaufit}.
The final velocity field and dispersion maps are shown Figure 8(a,b) and 11(a,b) for Sextans A and B respectively. 
 
\begin{table}
\scriptsize
\caption{\small Parameters of the KAT-7 Observations.}
\begin{minipage}{\textwidth}
\begin{tabular}{l@{\hspace{0.4cm}}c@{\hspace{0.4cm}}}   
\hline

Parameter & Sextans A $\&$ B \\
   
\hline \hline  
Start of observations & December 2014 \\
End of observations & June 2015\\
Total integration & 60 $\&$ 51 hours \\
FWHM of primary beam & 1.27$^{\circ}$\\
Total bandwidth & 12.5 MHz \\
Channel width (2 $\times$ 0.64 km.s$^{-1}$) & 1.28 km.s$^{-1}$\\
Number of channels (4096/2)&2048 \\
Map gridding &  30$^{\prime \prime}$ by 30$^{\prime \prime}$\\
Map size & 256 by 256\\
Flux/bandpass calibrator& 0407+658\\
Phase calibrator & 0941-080 \\
\hline    
\multicolumn{2}{@{} p{8.5 cm} @{}}{\footnotesize{\hspace{1cm} Robust = 0 weighting function}}\\\\
FWHM of synthesized beam & 228$^{\prime \prime} \times$ 201$^{\prime \prime}$ $\&$ 261$^{\prime \prime} \times$ 191$^{\prime \prime}$  \\
RMS noise & 4.5 $\&$ 5.0 mJy/beam\\
Column density limit&\\
(3$\sigma$ over 16 km.s$^{-1}$) & 5.8$\times$10$^{18}$ cm$^{-2}$ $\&$ 5.4$\times$10$^{18}$cm$^{-2}$ \\
\hline    

\label{coords_table}
 
\end{tabular}   

\end{minipage}
\end{table}  

\section{H{\sc i} Distribution}
\subsection{Sextans A}
The global H\textsc{i} profile of Sextans A is given in Figure 1. Plotted for comparison is the H\textsc{i} global profile from the VLA LITTLE THINGS data \citep{2012AJ....144..134H}. The mid-point velocity at the 50$\%$ level is 324 $\pm$ 2 km.s$^{-1}$, identical to the previous measurements \citet{1981A&A...102..134H,2002AJ....123.1476W,2004MNRAS.351..333B,2011AJ....142..173H}. The 50$\%$ line width is $\Delta V$ = 45 $\pm$ 2 km.s$^{-1}$. This is close to the value of  $\Delta V$ = 46 km.s$^{-1}$ derived by \citet{2004MNRAS.351..333B} but smaller than the value of  $\Delta V$ = 55 km.s$^{-1}$ derived by \citet{2002AJ....123.1476W}. An integrated flux of 181 $\pm$ 2 Jy km.s$^{-1}$ was measured, yielding a total H\textsc{i} mass of (7.3 $\pm$ 0.07) $\times$10$^{7}$ M$_{\odot}$ at the adopted distance of 1.3 Mpc. This is similar to the Parkes single dish results of \citet{2004MNRAS.351..333B} and in agreement within the error to their H\textsc{i} mass of 6.7 $\pm$ 0.5 $\times10^{7}$ M$_{\odot}$. 
The GBT single dish observations \citep{2011AJ....142..173H} reported an H\textsc{i} mass of 8.2$ \times 10^{7}$ M$_{\odot}$, which is 11$\%$ more compared to our derived H\textsc{i} mass. The GBT observations report $\sim$ 10$\%$ uncertainty on the flux due to calibration. Taking this into account brings the KAT-7 derived mass in agreement with the GBT calculated mass. This implies that KAT-7 does not miss out any galaxy flux. The KAT-7 derived H\textsc{i} mass is 5$\%$ more than the mass of 6.9 $\times 10^{7}$M$_{\odot}$ detected by the VLA mosaic observations \citep{2002AJ....123.1476W} and in excellent agreement with the H\textsc{i} mass of 7.1 $\times 10^{7}$ M$_{\odot}$ \citep{2012AJ....144..134H}. 

 \begin{figure}
  \includegraphics[width = \columnwidth]{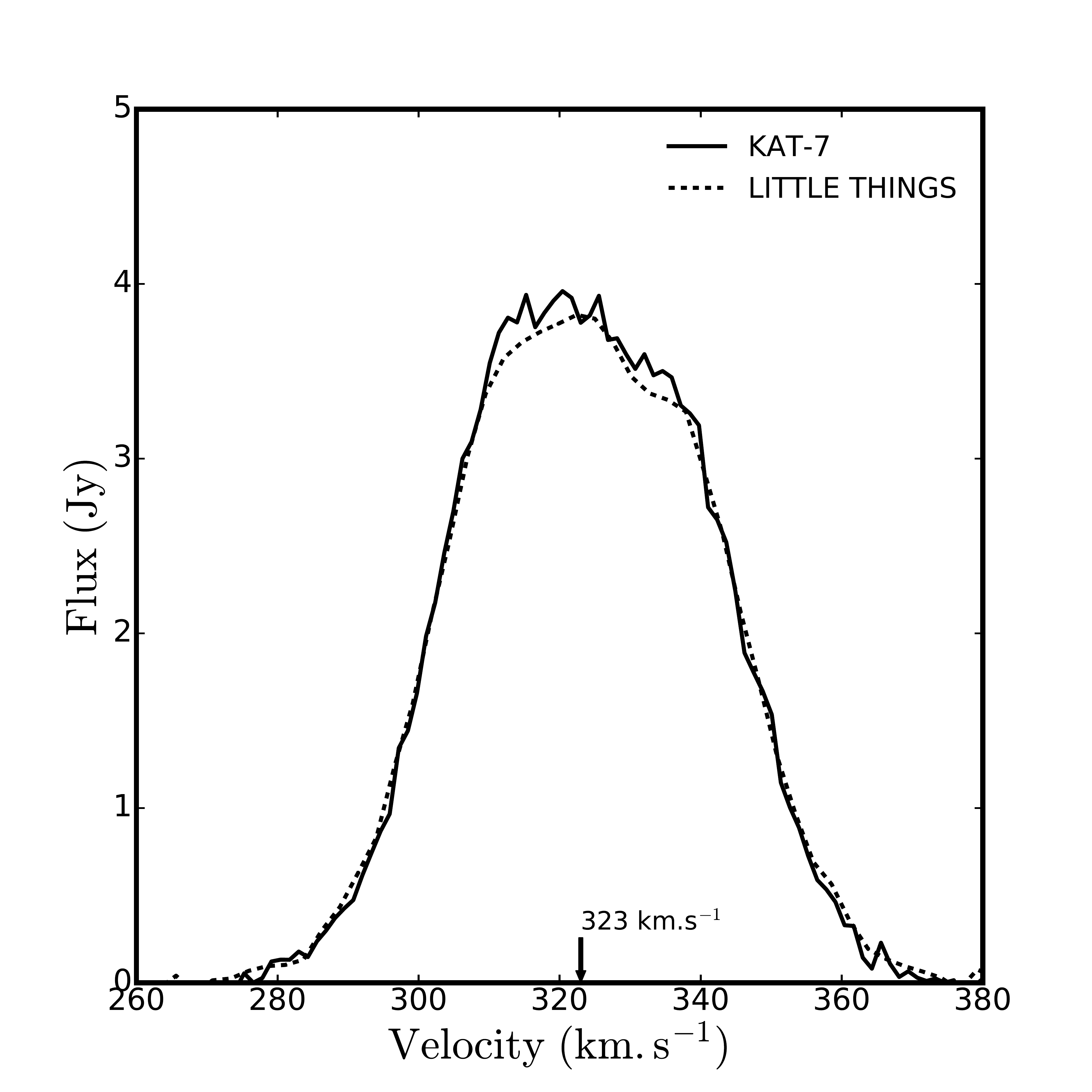}
  \caption{Global H\textsc{i} line profile of Sextans A from the KAT-7 primary beam corrected data cube, (black solid line) in comparison to the VLA LITTLE THINGS H\textsc{i} \citep{2012AJ....144..134H} global profile, (black dash-dotted line). The mid-point velocities of 323 km.s$^{-1}$ is indicated.} 
  \label{fig_speczvsphotz} 
\end{figure}
Figure 2 shows the integrated column density map of Sextans A superposed on the DSS image. The H\textsc{i} distribution is well resolved by the KAT-7 beam. The 3$\sigma$ column density limit in Figure 2 is 5.8$ \times 10^{18}$ cm$^{-2}$. This is higher than the 3$\sigma$ limit of 2 $\times 10^{18}$ cm$^{-2}$ of \citet{2011AJ....142..173H} and lower than the 7.5$ \times 10^{18}$ cm$^{-2}$ of \citet{2002AJ....123.1476W}. At the 3$\sigma$ column density level, we measure an H\textsc{i} diameter of 30$^{\prime}$. Comparing with the literature, we measure an H\textsc{i} diameter of 23$^{\prime}$ at 10$^{19}$ cm$^{-2}$ similar to an H\textsc{i} diameter of 22$^{\prime}$ of \citet{2011AJ....142..173H}.

 Figure 3 shows the azimuthally averaged KAT-7 radial H\textsc{i} density profile of Sextans A. Plotted for comparison is the H\textsc{i} radial profile from the VLA observations of \citet{2012AJ....144..134H}. Both profiles were derived using the GIPSY task \textsc{ellint} by applying the tilted ring kinematics parameters described in Section 5. It can be seen that the large beam of KAT-7 has averaged out the depression and peak in the center. 

\begin{figure}
  \includegraphics[width = \columnwidth]{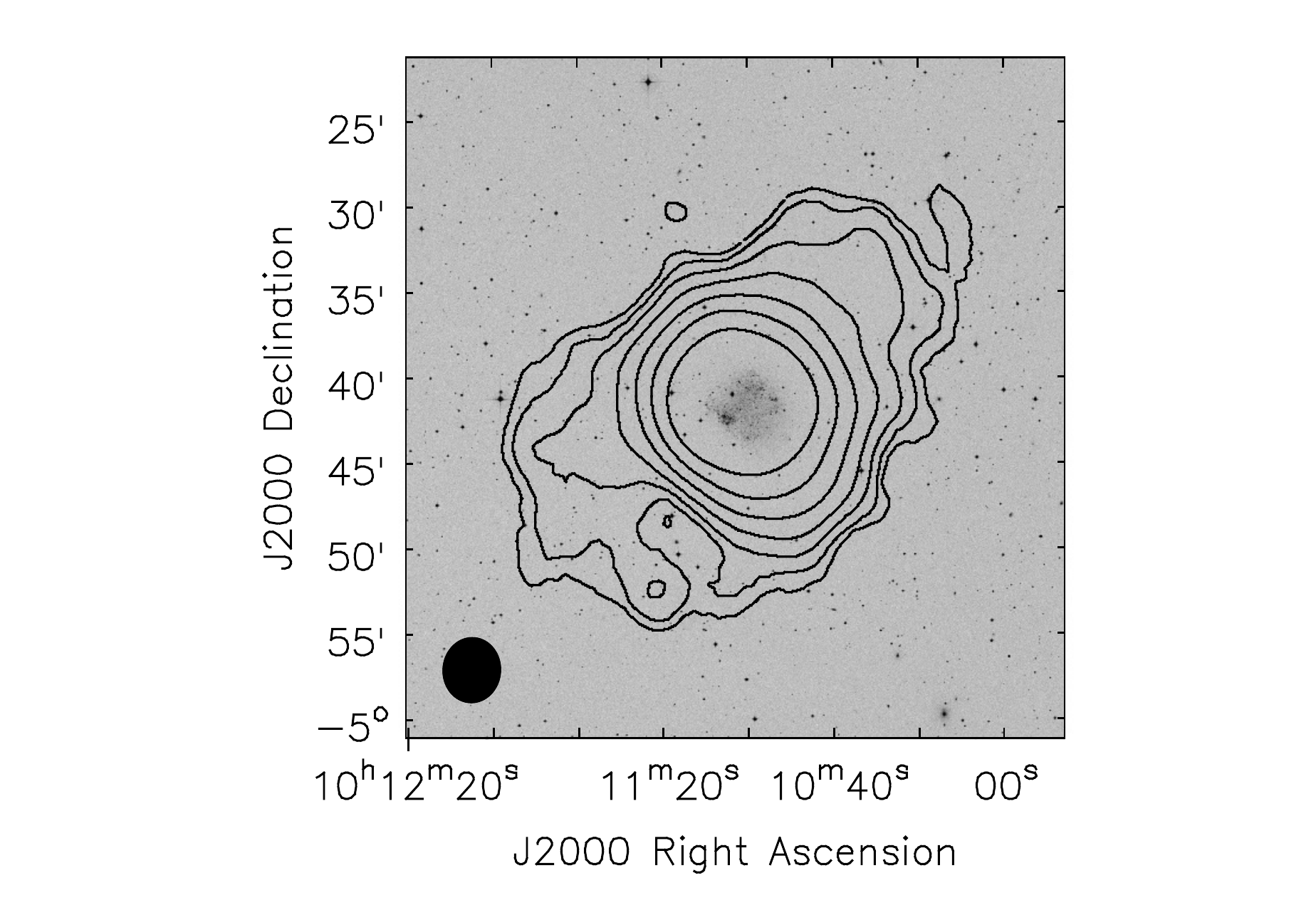}
  \caption{Integrated H\textsc{i} column density map of Sextans A superposed on a DSS image. The contours are 5.8, 11.6, 23.2, 46.4, 92.8, 185.0, and 370 $\times$ 10$^{18}$ cm$^{-2}$. The synthesized beam is shown in the bottom left corner.} 
  \label{fig_speczvsphotz} 
\end{figure}

 \begin{figure}
  \includegraphics[width = \columnwidth]{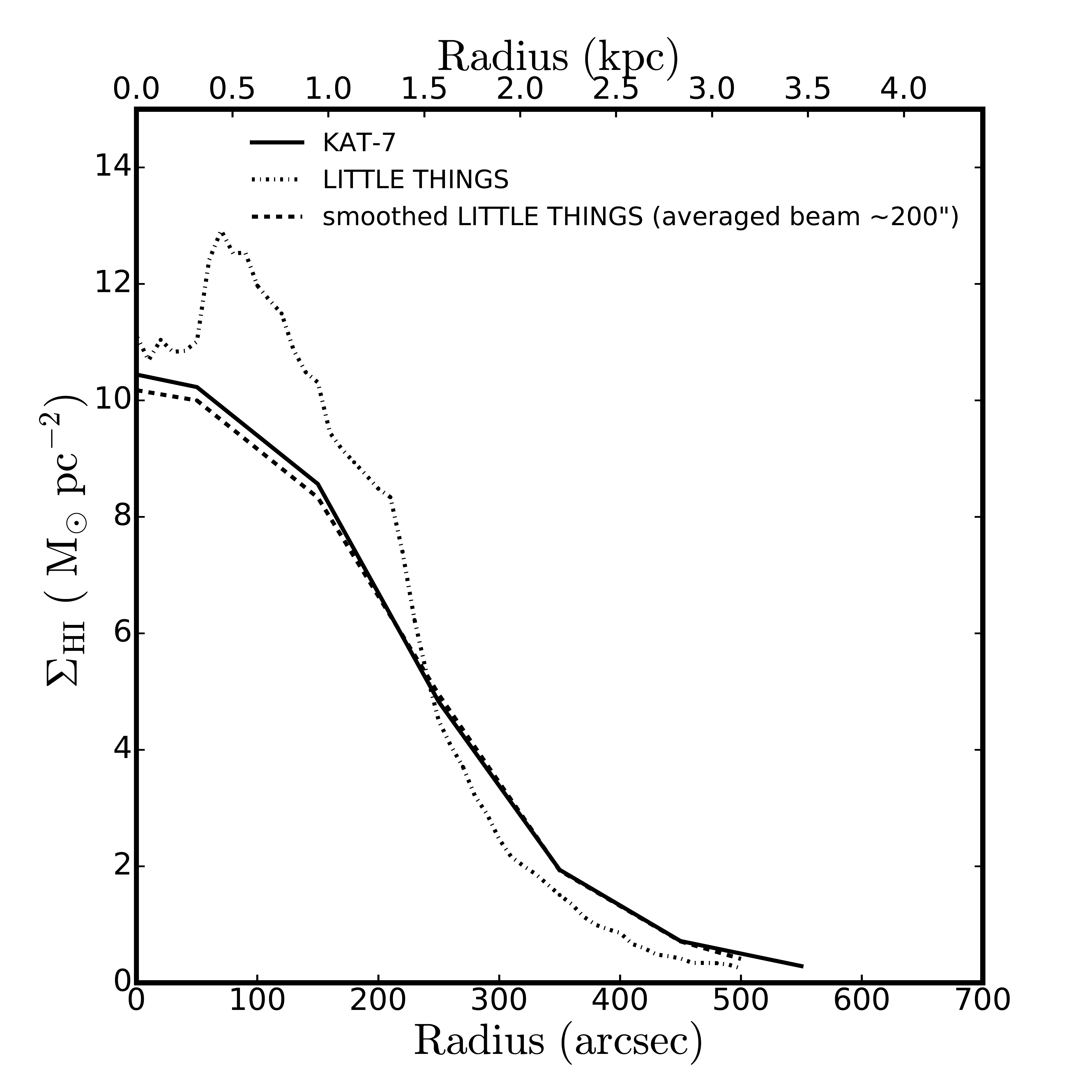}
  \caption{KAT-7 H\textsc{i} radial profile of Sextans A (black solid line) compared to the VLA LITTLE THINGS \citep{2012AJ....144..134H}. The dashed line shows the LITTLE THINGS radial profile smoothed to the KAT-7 spatial resolution while the dash-dotted lines shows LITTLE THINGS radial profile at VLA full spatial resolution. The surface densities are multiplied by a factor 1.4 to take into account helium and other metals.} 
  \label{fig_speczvsphotz} 
\end{figure}

\subsection{Sextans B}
The global H\textsc{i} spectrum of Sextans B is given in Figure 4. This is compared to the LITTLE THINGS derived global profile. The mid-point velocity at 50$\%$ level is 301 $\pm$ 1.4 Jy. km.s$^{-1}$. The 50$\%$ line width is $\Delta$V = 40.8 $\pm$ 1.84 km.s$^{-1}$. An integrated flux of 105$\pm$ 1.4 Jy.km.s$^{-1}$ is measured, giving a total H\textsc{i} mass of 4.2 $\times 10^{7}$ M$_{\odot}$. Using the same distance, the KAT-7 observations measure $\sim$ 15$\%$ more H\textsc{i} mass than the VLA-ANGST \citep{2012AJ....144..123O} and is in agreement with the value of 3.9 $\times 10^{7}$ M$_{\odot}$ derived from the VLA LITTLE THINGS \citep{2012AJ....144..134H}.   

Figure 5 shows the H\textsc{i} column density map of Sextans B superposed on a DSS optical image. The H\textsc{i} distribution is well resolved by the KAT-7 beam. The lowest contour is 5.4 $\times 10^{18}$ cm$^{2}$, which corresponds to the 3$\sigma$ limit calculated over 16 km.s$^{-1}$. At the lowest contour, the H\textsc{i} diameter is 20$^{\prime}$. The azimuthally averaged H\textsc{i} densities of Sextans B are shown in Figure 6. This is plotted together with the derived H\textsc{i} profile from the LITTLE THINGS data \citep{2012AJ....144..134H}. Again the beam smearing in the KAT-7 data has reduced the peak seen at high resolution.

\begin{figure}
  \includegraphics[width = \columnwidth]{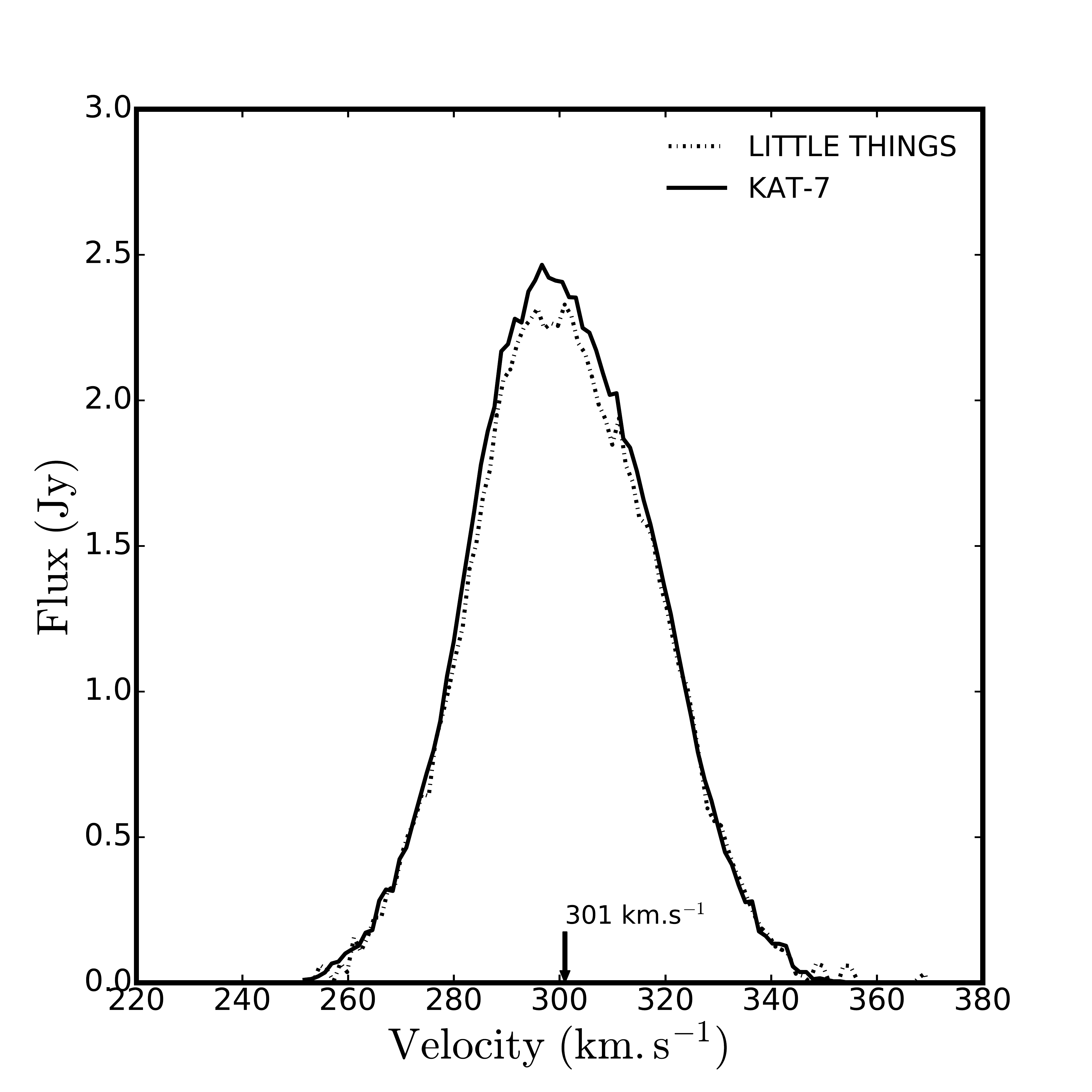}
  \caption{Global H\textsc{i} line profile of Sextans B from the KAT-7 primary beam corrected data cube (black solid line) compared to the VLA LITTLE THINGS H\textsc{i} \citep{2012AJ....144..134H} global profile (black dash-dot line). The mid-point velocity of 301 km.s$^{-1}$ is indicated.} 
  \label{fig_speczvsphotz} 
\end{figure}

\begin{figure}
  \includegraphics[width = \columnwidth]{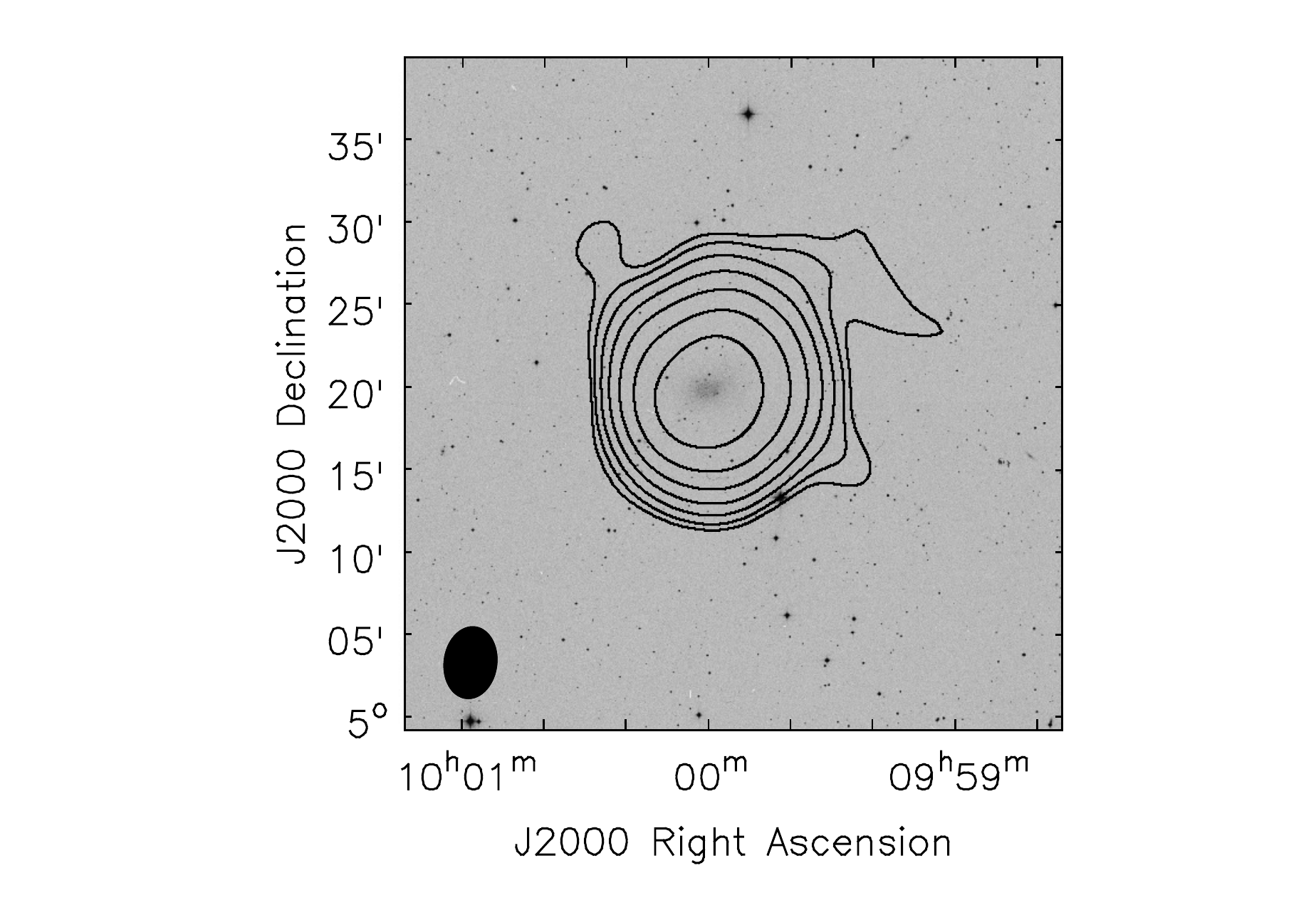}
  \caption{Integrated H\textsc{i} column density map of Sextans B superposed on a DSS image. The contours are 5.4, 10.8, 21.6, 43.2, 86.4, 172.8, and 345.6 $\times$ 10$^{18}$ cm$^{-2}$.  The synthesized beam is shown in the bottom left corner.} 
  \label{fig_speczvsphotz} 
\end{figure}

\begin{figure}
  \includegraphics[width = \columnwidth]{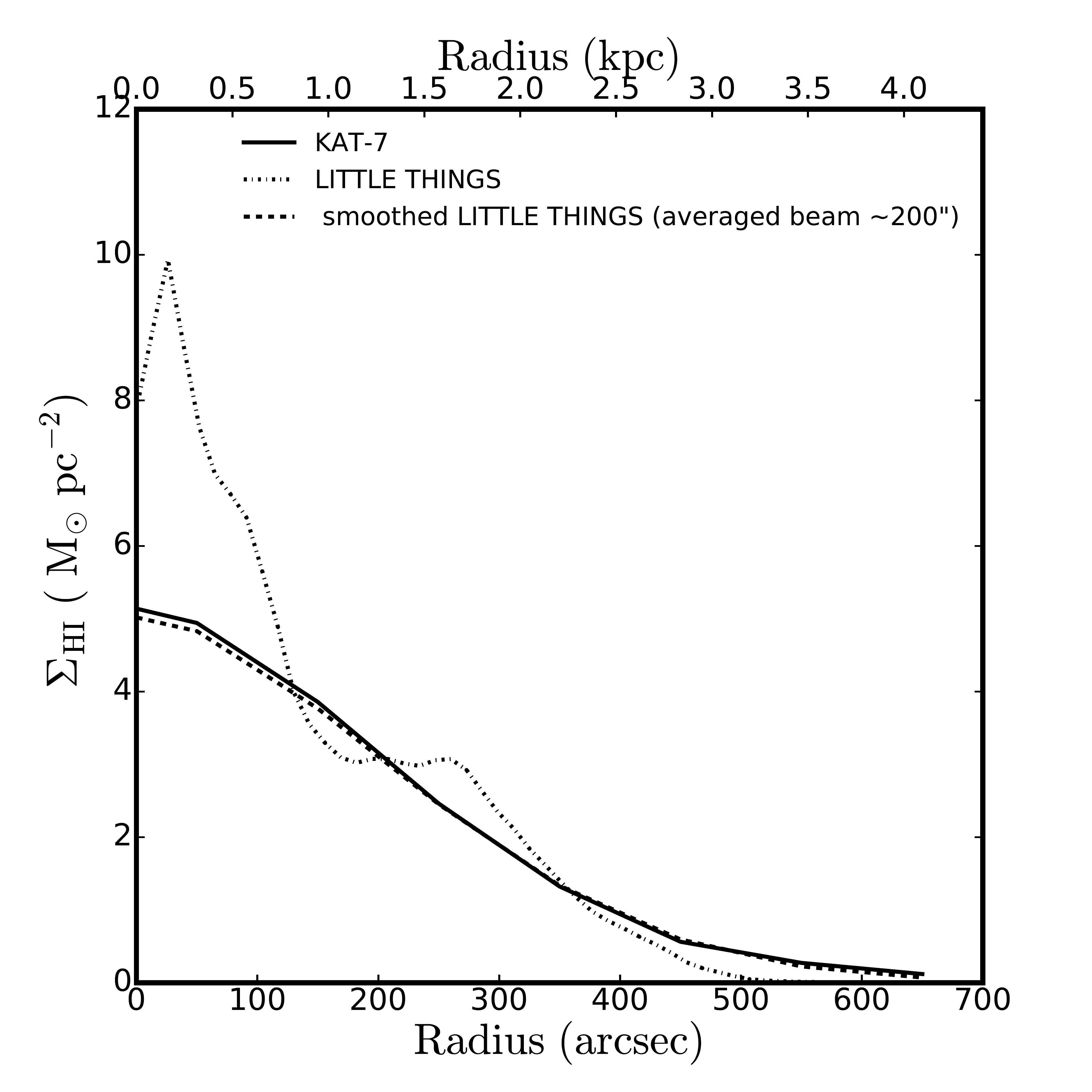}
  \caption{KAT-7 H\textsc{i} radial profile of Sextans B (black solid line) compared to the VLA LITTLE THINGS \citep{2012AJ....144..134H}. The dashed line shows the LITTLE THINGS radial profile smoothed to the KAT-7 spatial resolution while the dash-dotted lines shows LITTLE THINGS radial profile at VLA full spatial resolution. The surface densities are multiplied by a factor 1.4 to take into account helium and other metals.} 
  \label{fig_speczvsphotz} 
\end{figure}

\subsection{General remark on the H{\sc i} distribution}
It can be seen in Figure 1 and 4 that KAT-7 recovered a small fraction of flux missing from the VLA observations because of the lack of short spacings. However, the price to pay for having a lower spatial resolution is to wash out the high brightness features in the center, as can be seen in Figure 3 and 6. What is clear however is that the KAT-7 observations dismiss completely the large extents and fluxes derived for Sextans A from the Effelsberg observations. The KAT-7 total fluxes and H\textsc{i} extents are much closer to the ones derived using the GBT or the VLA (see Table 3). 
 
\begin{table}
\scriptsize

\caption{\small Comparison of the H\textsc{i} fluxes and extents of Sextans A and B.}
\begin{minipage}{\textwidth}
\begin{tabular}{l@{\hspace{0.2cm}}c@{\hspace{0.0001mm}}c@{\hspace{0.0001mm}}c@{\hspace{0.0001mm}}c@{\hspace{0.0001mm}}}   
\hline

   
\hline \hline  
&&Integrated flux (Jy.km.s$^{-1})$&& \\
\cline{2-3}
Galaxy &KAT-7$^{a}$ &Effelsberg$^{b}$&GBT$^{c}$& VLA$^{d}$\\ 
sextans A & 181 $\pm$ 2.0& 264& 206$\pm$ 20.6&178 $\pm$ 2.0\\ 
Sextans B& 105 $\pm$ 1.4& 116& ....&98 $\pm$ 1.2\\ \hline
&&H\textsc{i} extent ($^{\prime}$) at 10$^{19}$ atoms.cm$^{-2}$&& \\
\cline{2-3} 
& 23.0& 54.0& 22.1&22.0\\ 
& 14.2& 13.0&....&12.8\\ \hline

\hline  

\multicolumn{5}{@{} p{9.0 cm} @{}}{\footnotesize{\textbf{Notes.} Ref (a) This work; (b)\citet{1981A&A...102..134H}; (c) \citet{2011AJ....142..173H}; (d)\citet{2012AJ....144..134H} }}
\label{coords_table}

\end{tabular}   

\end{minipage}
\end{table}  
\section{H\textsc{i} kinematics}
\subsection{Tilted ring model}
We fit a tilted-ring model to the velocity fields using ROTCUR in GIPSY \citep{1989A&A...223...47B} to derive the parameters that best describe the observed velocity fields.

\subsection{Deriving the rotation curve}

Following \citet{2017arXiv170809447N} we derive the rotation curves of Sextans A and B using a spacing and width of 100$^{\prime \prime}$. The errors on V$_{rot}$ were derived using Equation 2 \citep{2013AJ....146...48C}.
\begin{equation}
\Delta V = \sqrt{\sigma^{2}(V) + (\frac{|V_{app} - V_{rec}|}{2})^2}
\end{equation}
We correct the rotational velocities for the asymmetric drift as the dynamical support by random motions of the gas disk is significant. Following the method described in \citet{2000AJ....120.3027C}, we correct for the asymmetric drift as follows:
\begin{equation}
V_{c}^{2} = V_{0}^{2} - 2\sigma \frac{\delta \sigma}{\delta \ln R } - \sigma^{2} \frac{\delta \ln \Sigma}{\delta \ln R}
\end{equation}
where $V_{c}$ is the corrected velocity, V$_{0}$ is the observed one, $\sigma$ is the velocity dispersion and $\Sigma$ is the gas density. 

\subsection{Sextans A rotation curve results}
Results of the tilted ring model fitted to the first order gaussian velocity field for Sextans A are presented in Figure 7 and 8. The first and second panel of Figure 7 shows the behavior of the inclination and position angle when the parameters are left free to vary while the green solid lines show the fitted parameters used to derive the final rotation curves for the approaching, receding, and both sides, shown in the third panel of Figure 7. We derive the rotation curve out to 550$^{\prime \prime}$, which is $\sim$ 3.5 kpc. The rotation curve rises as V(R) $\propto$ R out to $\sim$ 250$^{\prime \prime}$. Beyond this radius, the rotation curve is seen to decline down to 550$^{\prime \prime}$. This feature in the RC is not peculiar to dwarf galaxies such as Sextans A. The analysis of the rotation curve of GR 8 \citep{1990AJ.....99..178C} showed that the RC of the galaxy was declining in the outer regions (see also e.g. the case of the late-type spiral NGC 7793: \citet{2008AJ....135.2038D}). A systemic velocity of V$_{sys}$ = 324 $\pm$ 0.6 km.s$^{-1}$ is found, similar to the value obtained from the global profile. A mean $i$ and $P.A.$ of 34$^{\circ}$ and 86$^{\circ}$ are measured respectively. 

Figure 8 compares the observed velocity field (a) to the derived model velocity field (c), while (d) shows the residual map (observed velocity field - model velocity field). (b) shows the velocity dispersion map. We see no systemic large scale residuals, with most of them less than 5 km.s$^{-1}$. Only a few residuals $\sim$ 10 km.s$^{-1}$ 
are seen in the low S/N outer parts. Figure 9 shows the comparison between our KAT-7 rotation curve and the VLA LITTLE THINGS curve (full resolution and smoothed to KAT-7 spatial resolution)\citep{2012AJ....144..134H}. The rotation curves are in agreement within errors.

\begin{figure*}
\centering
\resizebox{1.0\hsize}{!}{\rotatebox{0}{\includegraphics{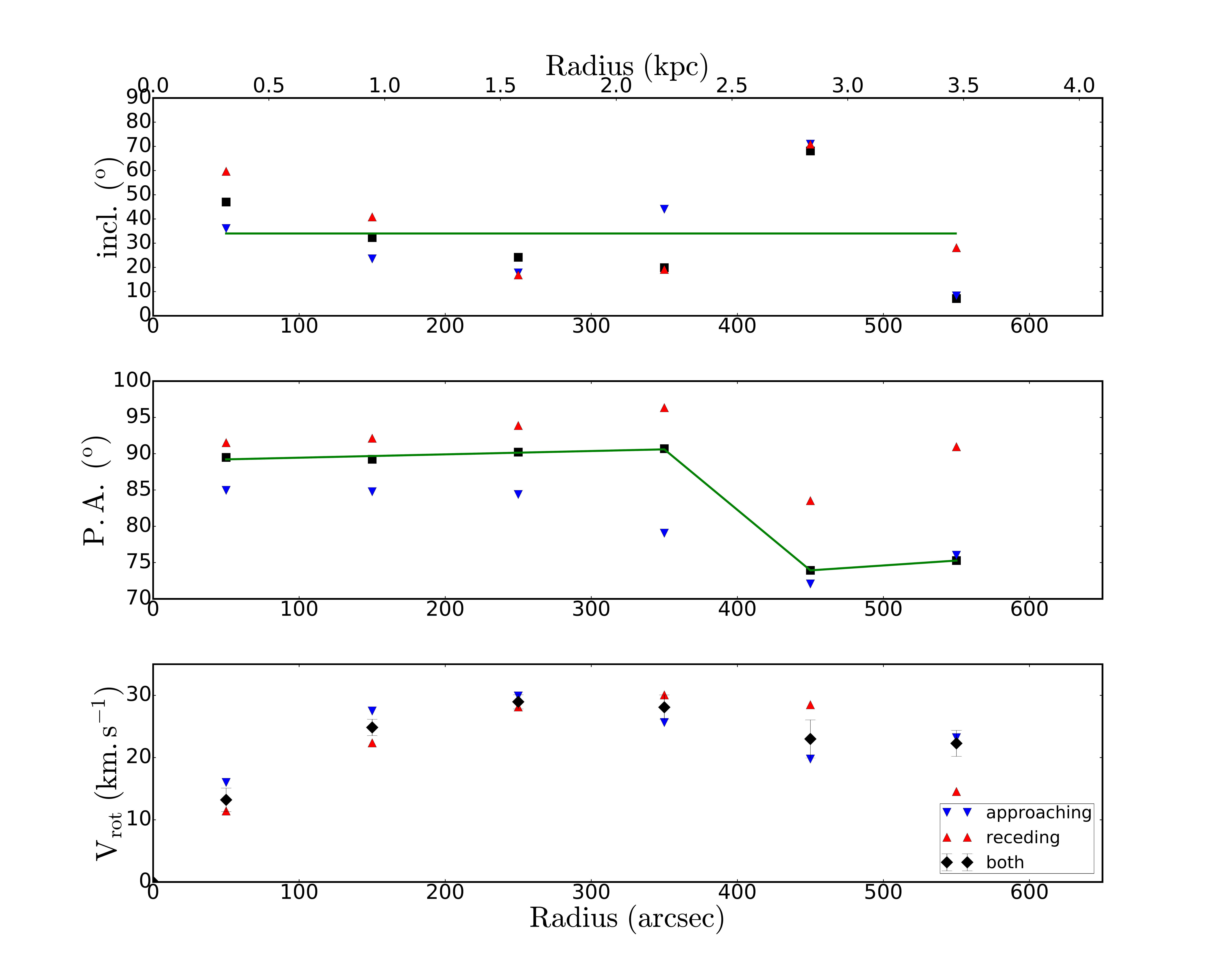}}}
\caption{Results of the tilted ring fits for Sextans A. The blue and red triangles in the middle and top panels show the behavior of the PA and inclination as free parameters. The black squares show the radial variation of the PA and inclination derived for both sides. The green solid lines show the behavior of the PA and inclination fixed to the model used to derive the final rotation curve. In this case the PA is varying while the inclination is fixed to the mean value. For the bottom panel, the blue triangles represent the curve for the approaching side, the red triangles represents the curve for the receding side, and the black diamonds show the rotation curve derived from both sides.}
  \label{fig_speczvsphotz} 
\end{figure*}

\begin{figure*}
\centering
   \subcaptionbox{Observed velocity field map \label{fig3:b}}{\includegraphics[width=3.0in]{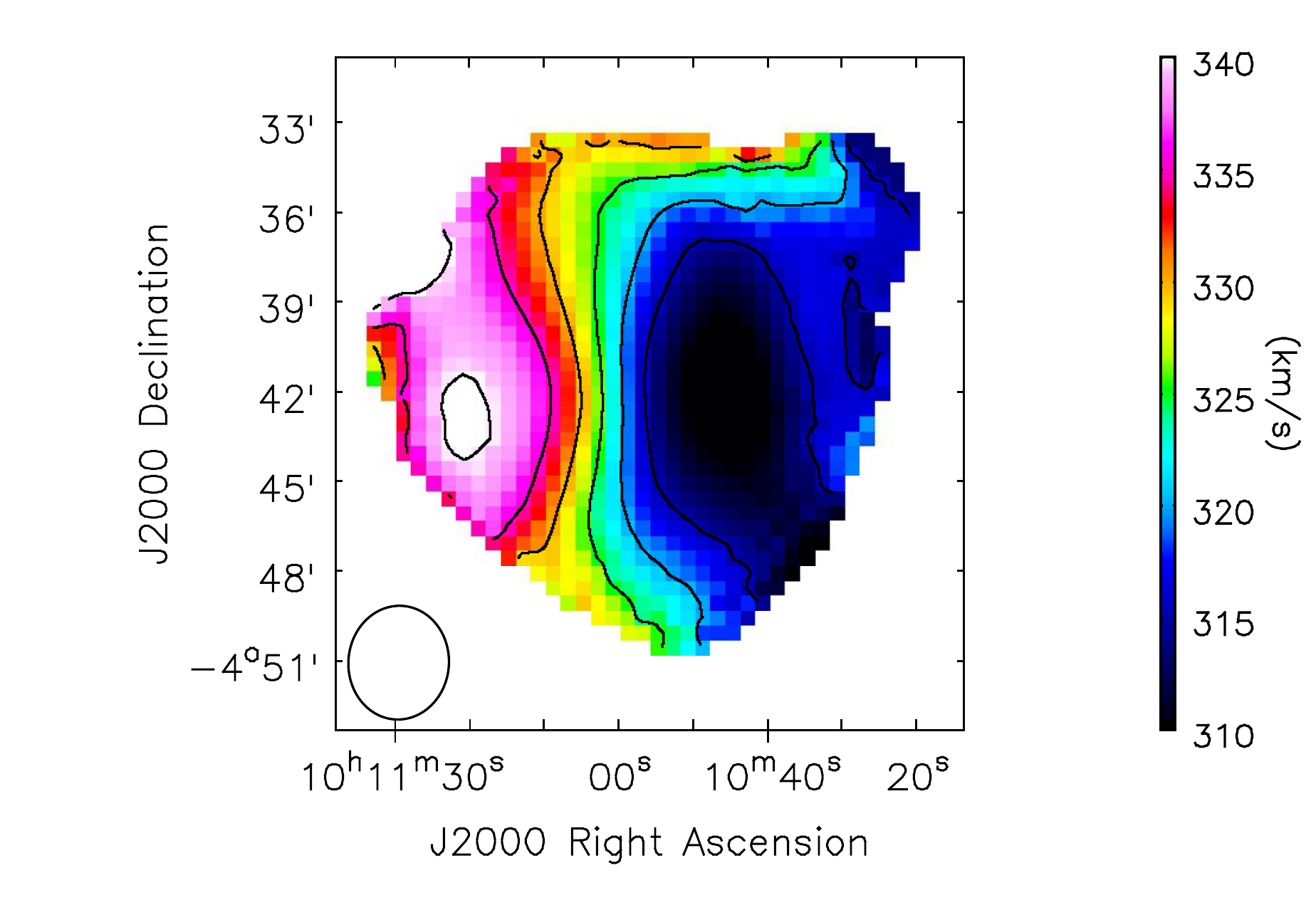}} \hspace{-1mm}  
   \subcaptionbox{Velocity dispersion map \label{fig3:a}}{\includegraphics[width=3.0in]{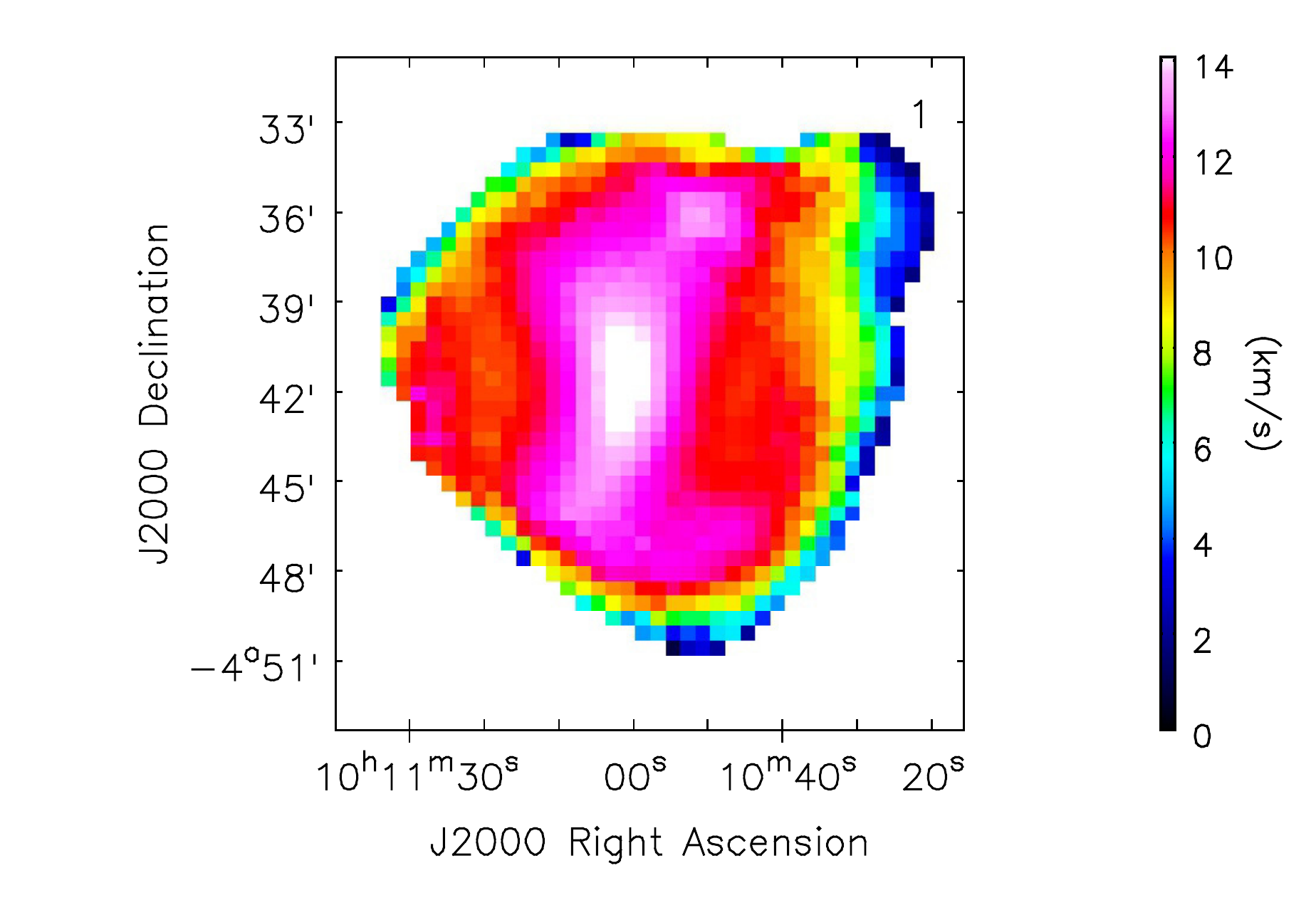}}\hspace{-1mm}\\
   \subcaptionbox{Model velocity field map\label{fig3:b}}{\includegraphics[width=3.0in]{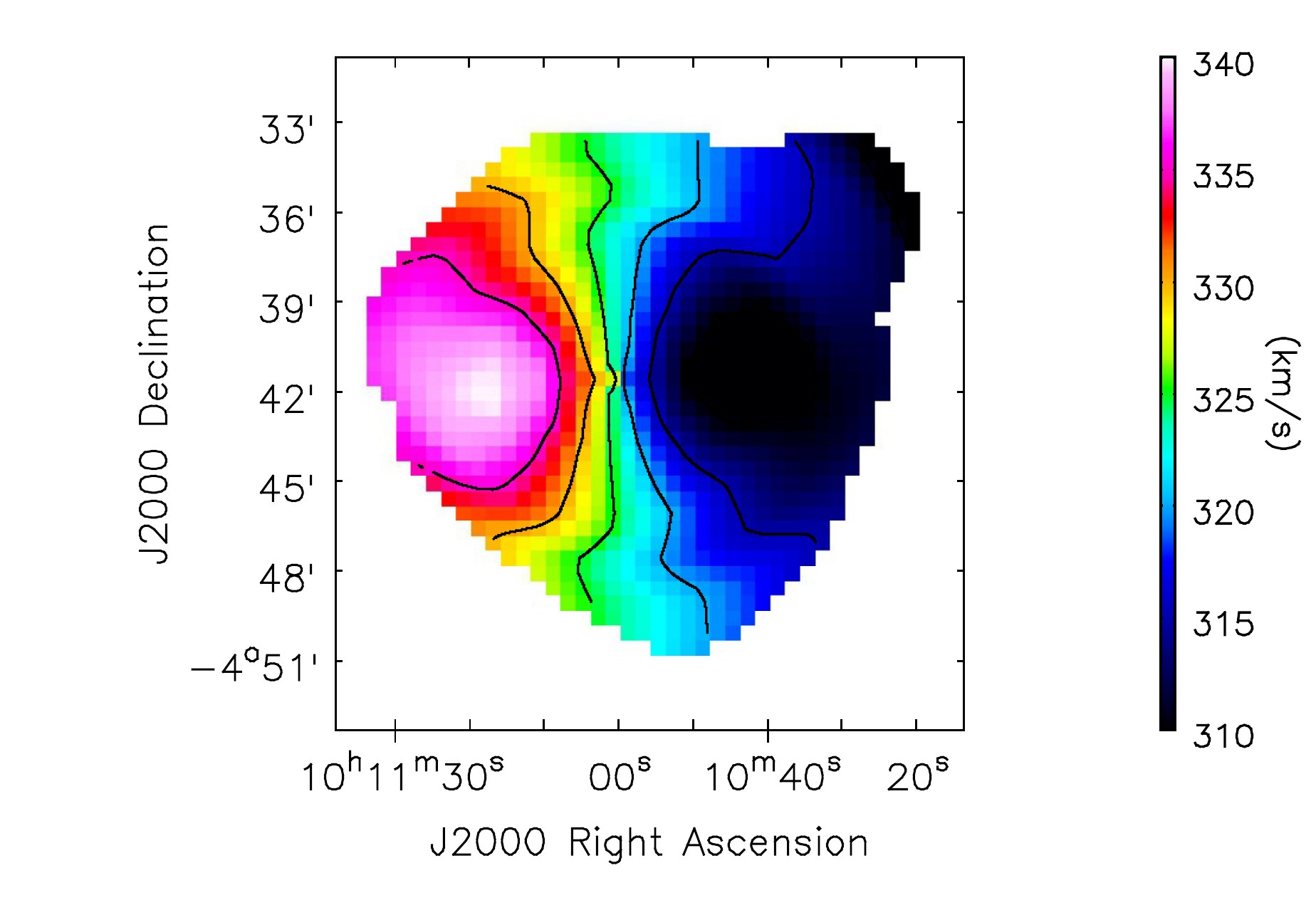}} \hspace{-1mm} 
   \subcaptionbox{Residual velocity field map\label{fig3:a}}{\includegraphics[width=3.0in]{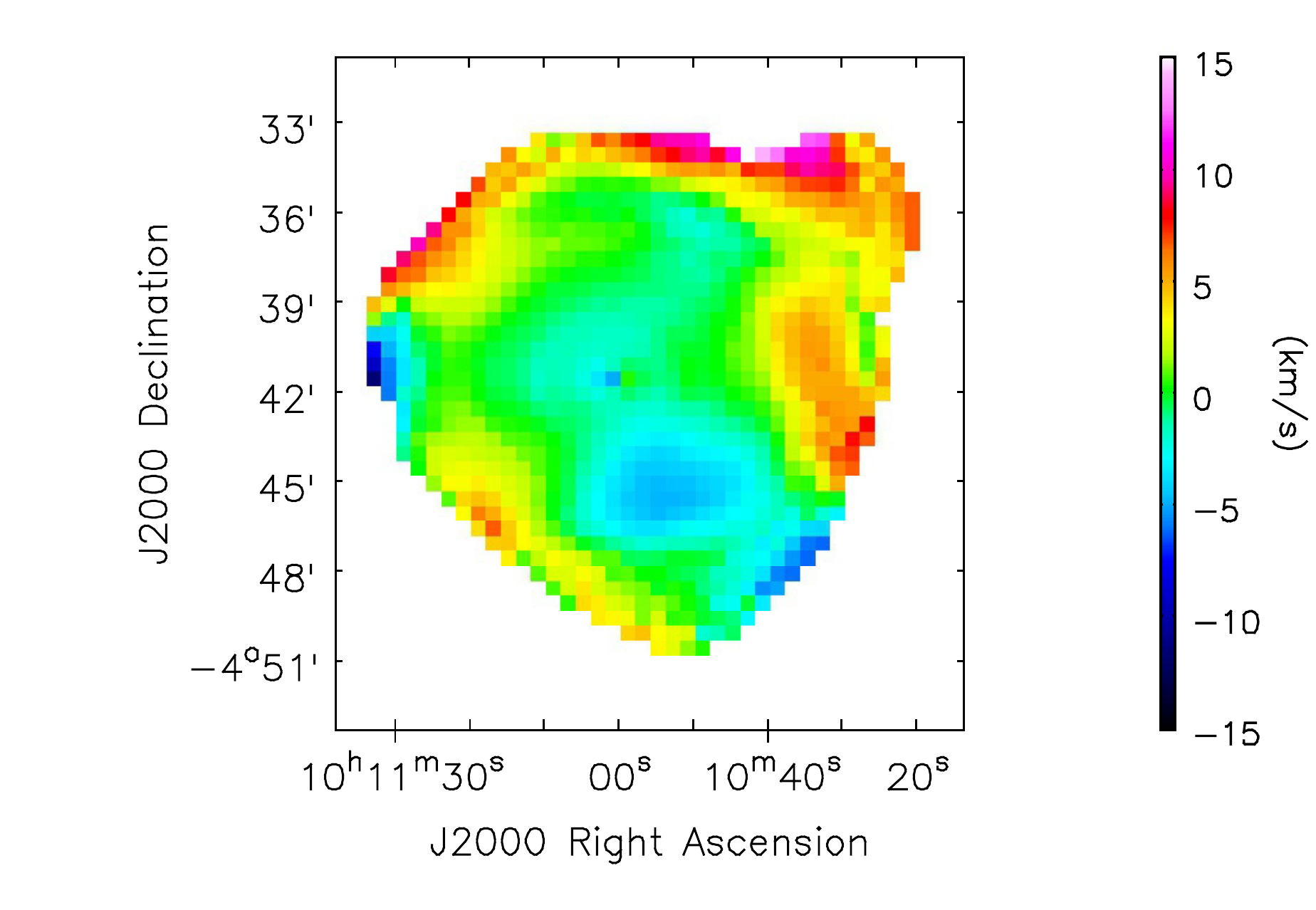}}\hspace{-1mm}\\%
   \caption{Maps of Sextans A: Observed velocity field map (a), velocity dispersion map (b), model velocity field (c), and residual map (d). The observed and model velocity field
   contours run from 310 to 340 km.s$^{-1}$ in steps of 5 km.s$^{-1}$. }  
   \end{figure*}

%
\begin{figure}
\centering
\resizebox{1.0\hsize}{!}{\rotatebox{0}{\includegraphics{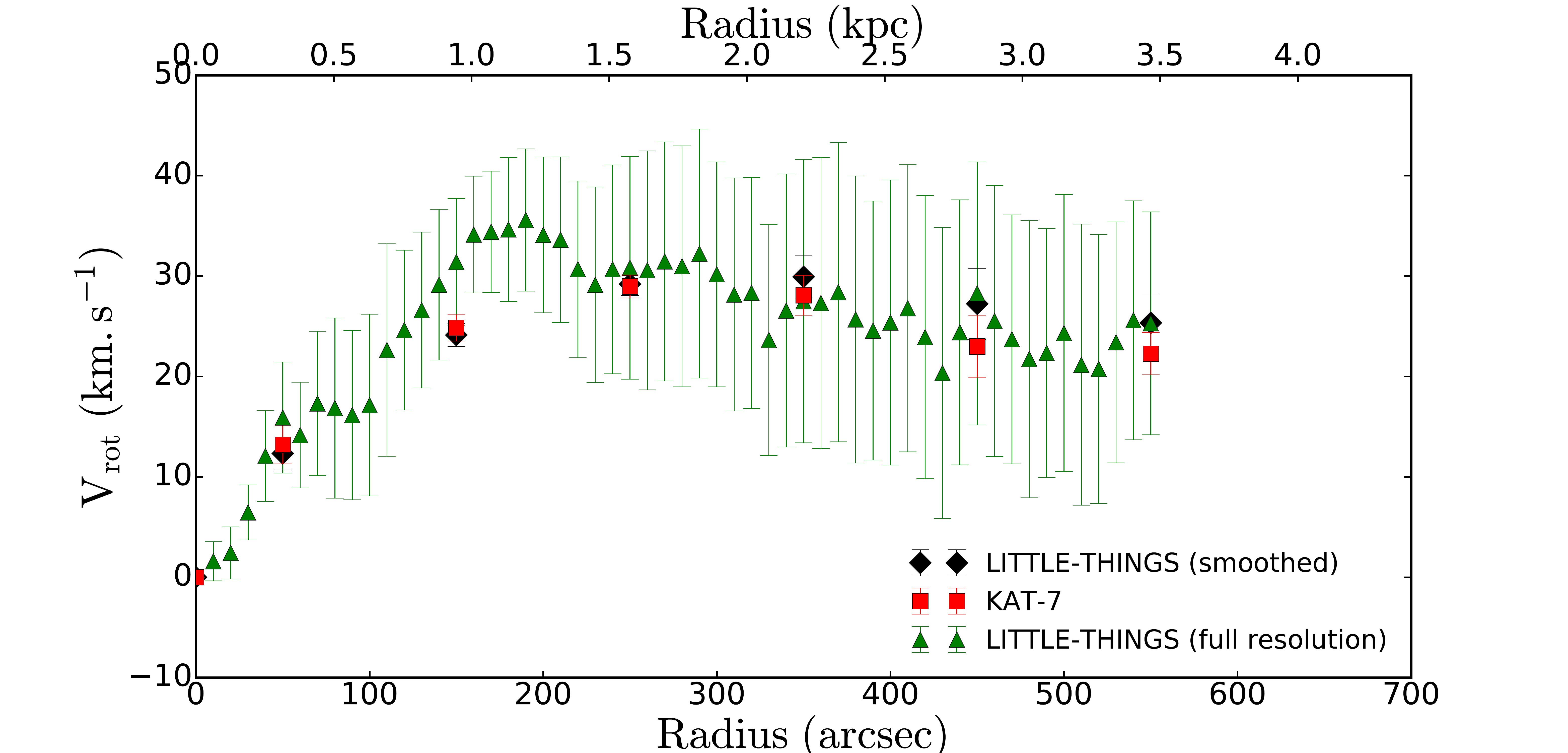}}}
\caption{Comparison of the KAT-7 and VLA LITTLE THINGS rotation curves of Sextans A. The red square show the KAT-7 rotation curve, the black triangles show the LITTLE THINGS RC smoothed to the KAT-7 spatial resolution, and the green triangles show the full resolution LITTLE THINGS RC.}
  \label{fig_speczvsphotz} 
\end{figure}

\begin{table}   
\caption{\small Radial variation of the H\textsc{i} surface densities $\Sigma_{g}$, the gas velocity dispersions $\sigma$, the observed rotation velocities $V_{0}$, the errors of the observed velocities $\Delta V$, and the asymmetric drift corrected rotation velocities $V_{c}$ for the KAT-7 data of Sextans A.}
\begin{minipage}{\textwidth}
\begin{tabular}{l@{\hspace{0.30cm}}c@{\hspace{0.30cm}}c@{\hspace{0.30cm}}c@{\hspace{0.30cm}}c@{\hspace{0.30cm}}c@{\hspace{0.30cm}}}   
\hline

Radius & $\Sigma_{g}$   &  $\sigma $ &  V$_{0}$ & $\Delta$ V  &  V$_{\text{{c}}}$  \\
arcsec & M$_{\odot}$pc$^{-2}$    &   km.s$^{-1}$   &    km.s$^{-1}$ &  km.s$^{-1}$ & km.s$^{-1}$ \\ \hline \hline
0.0&7.4&14.1&0.0&0.0&0.0\\
50&7.3&14.0&13.2&2.9&13.0\\
150&6.1&12.5&24.8&2.8&26.4\\
250&3.4&11.3&28.9&1.4&33.2\\
350&1.4&10.9&28.1&2.9&35.3\\
450&0.5&10.5&22.9&5.3&32.9\\
550&0.2&9.9&22.2&4.8&31.8\\

\hline    
\multicolumn{6}{@{} p{8.5 cm} @{}}{\footnotesize{\textbf{Notes.} Column (1) gives the radius, column (2) the surface densities, column (3) the velocity dispersion, column(4) the observed rotation velocities, column (5) the errors of those velocities, and column (6) the corrected velocities used for the mass models.}}
\label{coords_table}
 
\end{tabular}   

\end{minipage}
\end{table}  

\subsection{Sextans B rotation curve results}
The results of the tilted ring model fitting for Sextans B are shown in Figure 10 and 11. A constant inclination and a varying P.A. were used to derive the final rotation curve. A constant inclination was chosen as the variation of the inclination with radius was small. We measure the rotation curve out to 650$^{\prime \prime}$, which corresponds to $\sim$ 4 kpc. The rotation curve is seen to be rising out to $\sim$ 550$^{\prime \prime}$, with the last point showing a decline. A declining RC of Sextans B has been reported by \citet{2015AJ....149..180O}. A mean systemic velocity, V$_{sys}$ of 302 $\pm$ 0.9 km.s$^{-1}$ is found, similar to the value derived from the global profile. We find a mean $i$ and $P.A.$ of 49$^{\circ}$ and 56.6$^{\circ}$ respectively, values consistent with the literature \citet{2015AJ....149..180O}.

In Figure 11, we compare the observed velocity field (a) to the derived model velocity field (c), while (d) shows the residual map (observed velocity field - model velocity field). (c) shows the velocity dispersion map. It can be seen that the model is a very good representation of the velocity field with most residuals smaller than 5 km.s$^{-1}$. In Figure 12, the KAT-7 rotation curve of Sextans B is compared to the RC derived from the VLA LITTLE THINGS data \citep{2012AJ....144..134H} (full resolution and smoothed to KAT-7 spatial resolution). The rotation curves do not show a good agreement as for Sextans A, especially in the inner parts, where the effect of beam smearing of the KAT-7 observations is more important than in Sextans A.

\begin{figure*}
\centering
\resizebox{1.0\hsize}{!}{\rotatebox{0}{\includegraphics{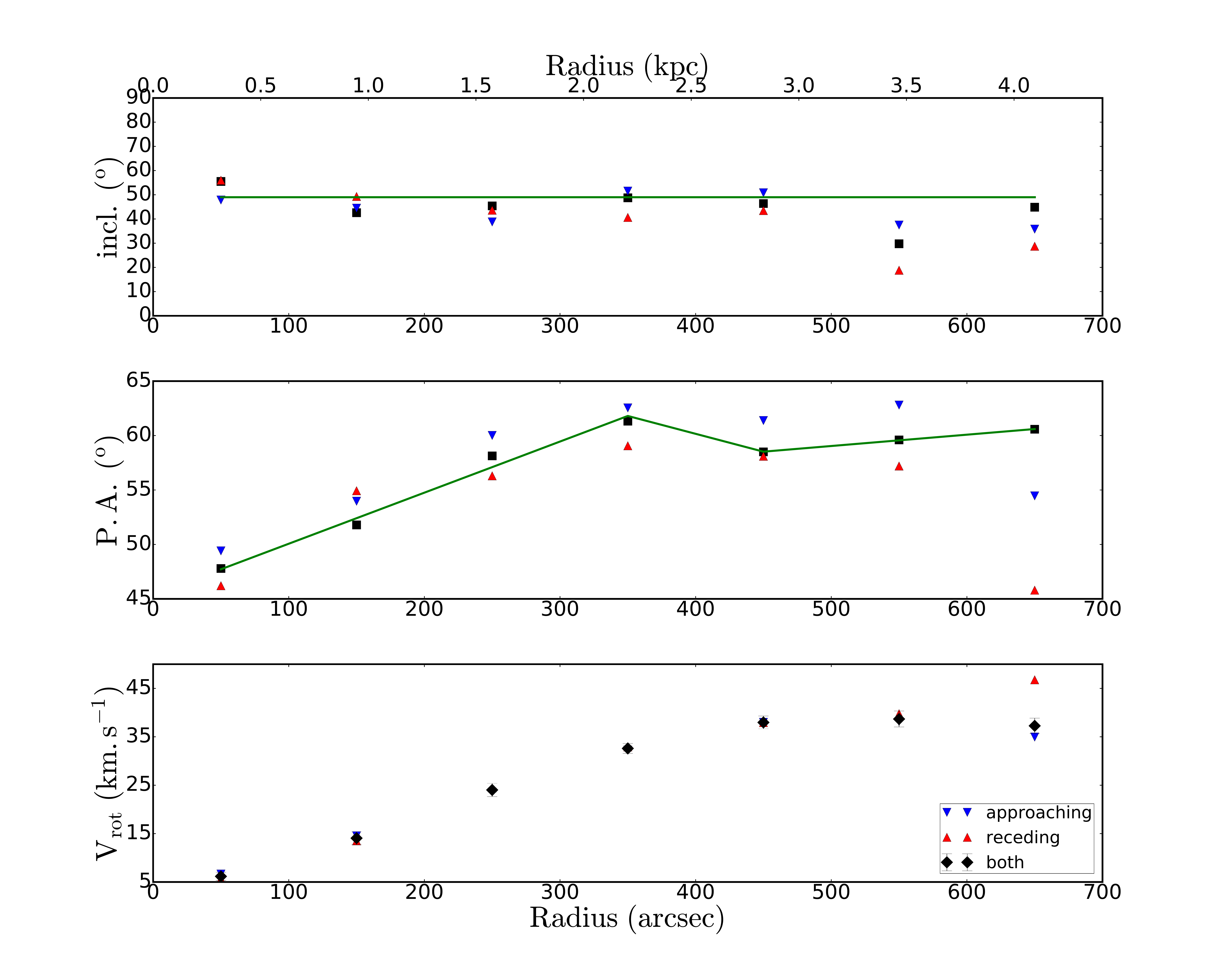}}}
\caption{Results of the tilted ring fits for Sextans B. The blue and red triangles in the middle and top panels show the behavior of the PA and inclination as free parameters. The black squares show the radial variation of the PA and inclination derived for both sides. The green solid lines show the behavior of the PA and inclination fixed to the model used to derive the final rotation curve. In this case the PA is varying while the inclination is fixed to the mean value. For the bottom panel, the blue triangles represent the curve for the approaching side, the red triangles represents the curve for the receding side, and the black diamonds show the rotation curve derived from both sides.}
  \label{fig_speczvsphotz} 
\end{figure*}

\begin{figure*}
\centering
   \subcaptionbox{Observed velocity field map\label{fig3:b}}{\includegraphics[width=3.0in]{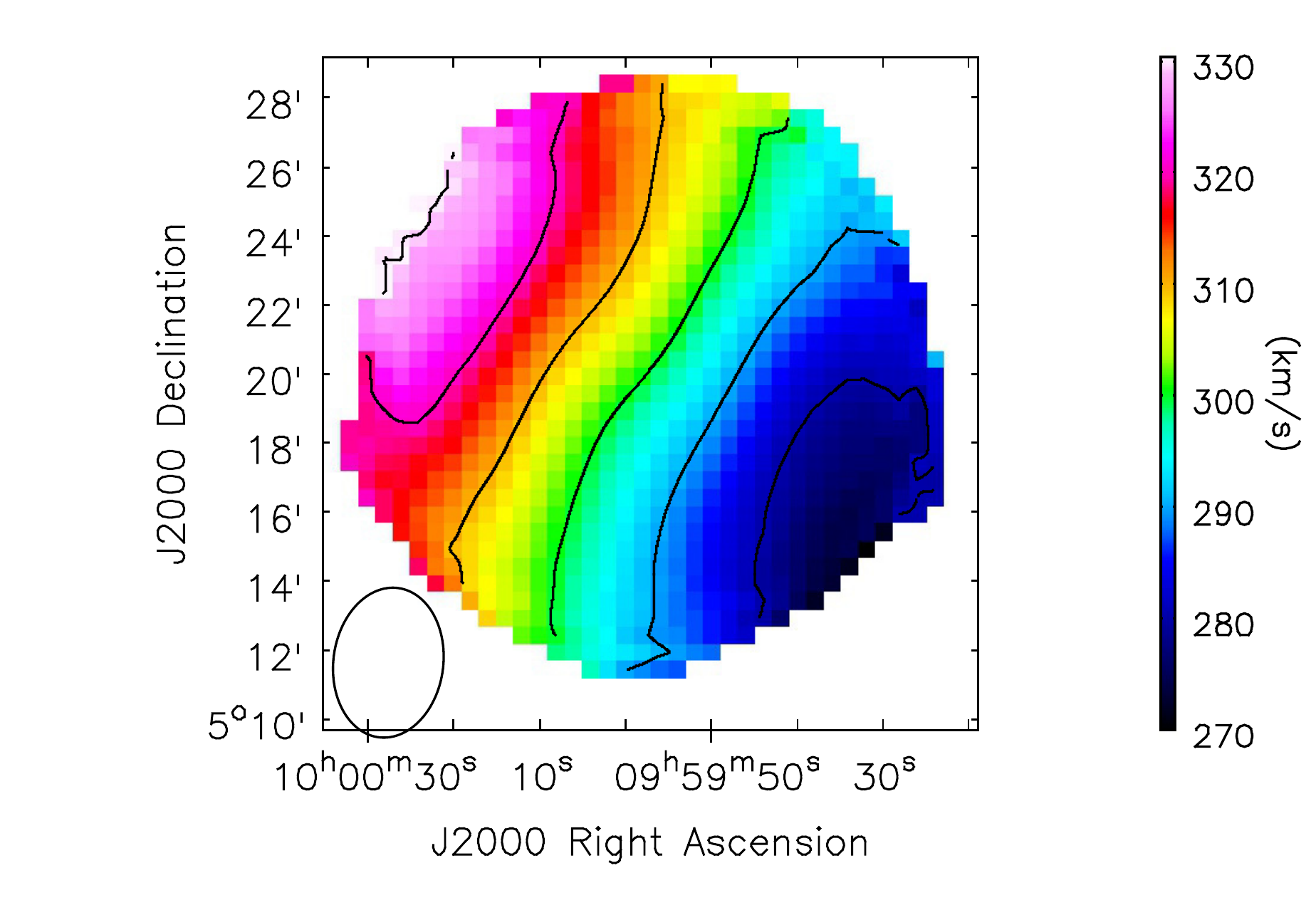}} \hspace{-1mm}  
   \subcaptionbox{Velocity dispersion map\label{fig3:a}}{\includegraphics[width=3.0in]{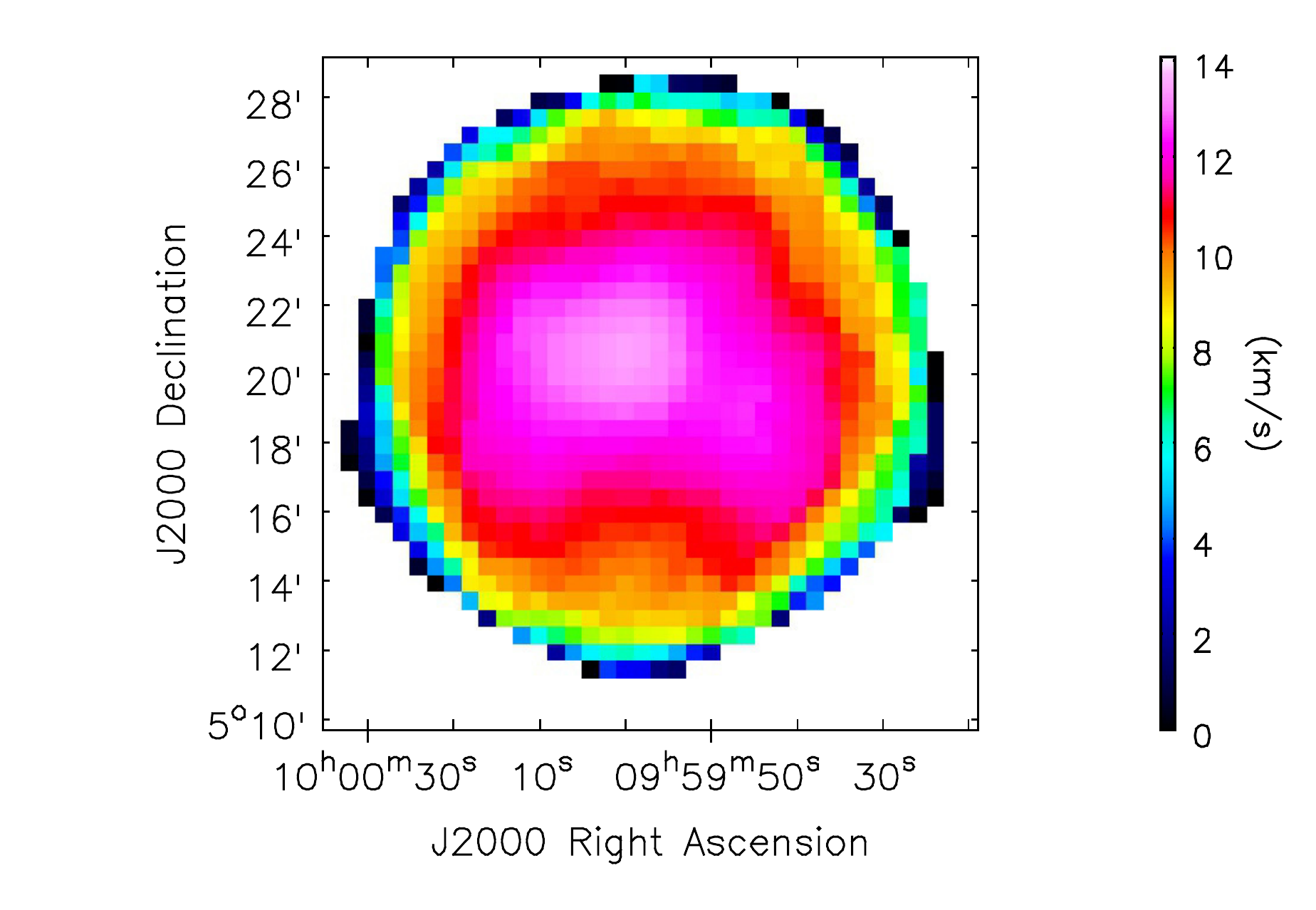}}\hspace{-1mm}\\%
   \subcaptionbox{Model velocity field map \label{fig3:b}}{\includegraphics[width=3.0in]{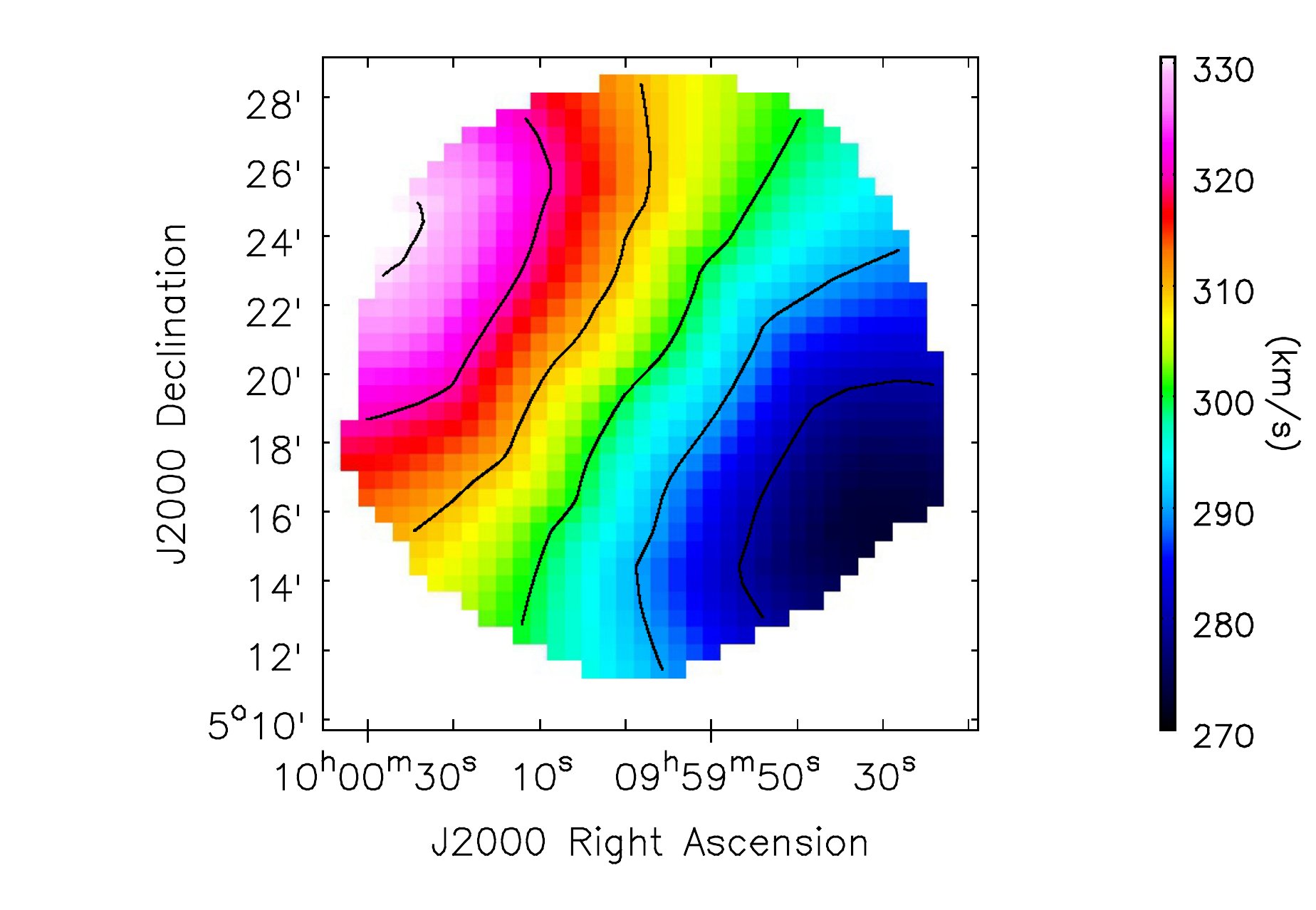}} \hspace{-1mm}  
   \subcaptionbox{Residual velocity field map\label{fig3:a}}{\includegraphics[width=3.0in]{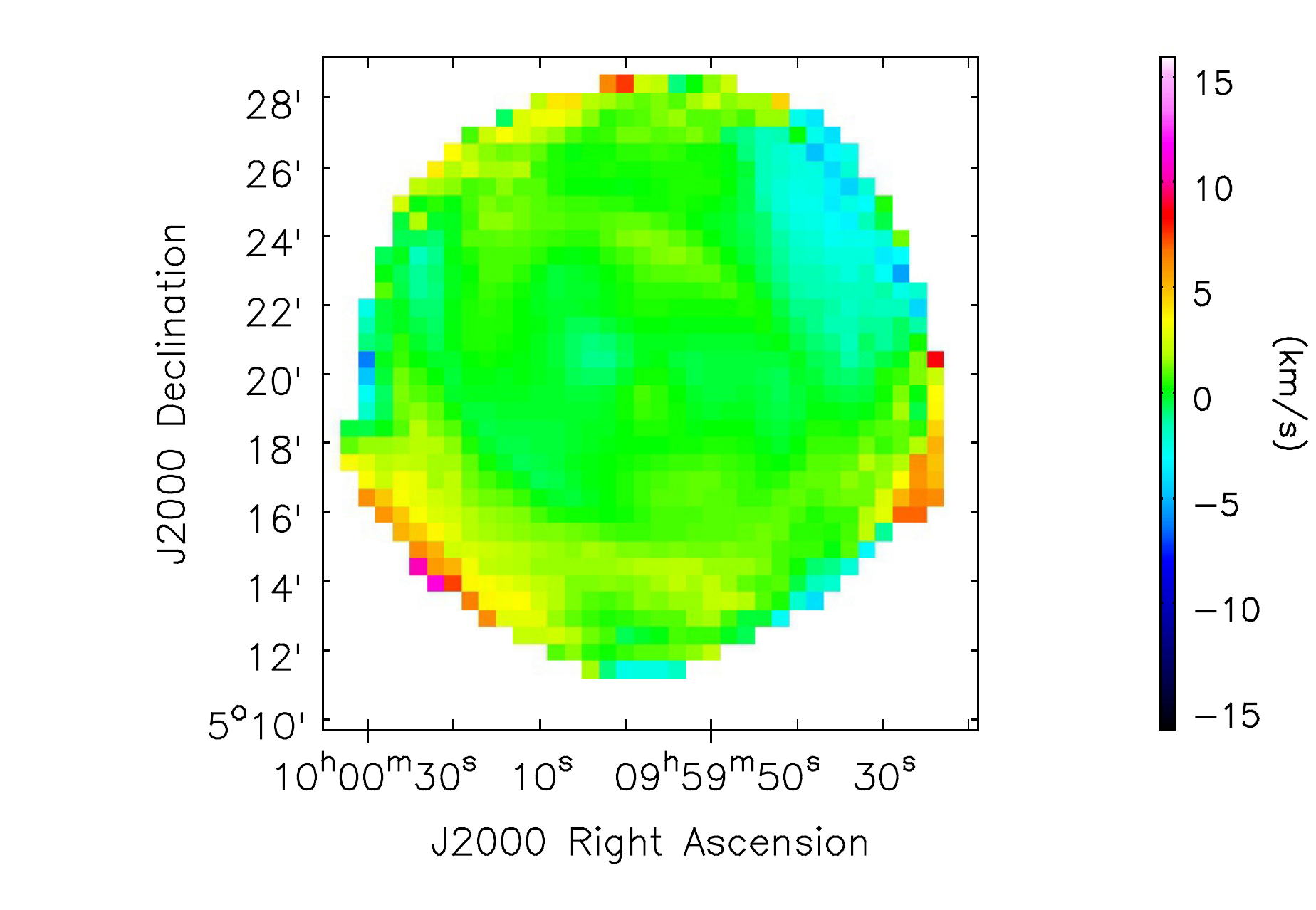}}\hspace{-1mm}\\%
   \caption{Maps of Sextans B: Observed velocity field map (a), velocity dispersion map (b), model velocity field (c), and residual map (d). The observed and model velocity field
   contours run from 280 to 340 km.s$^{-1}$ in steps of 10 km.s$^{-1}$. }  
   \end{figure*}


\begin{figure}
\centering
\resizebox{1.0\hsize}{!}{\rotatebox{0}{\includegraphics{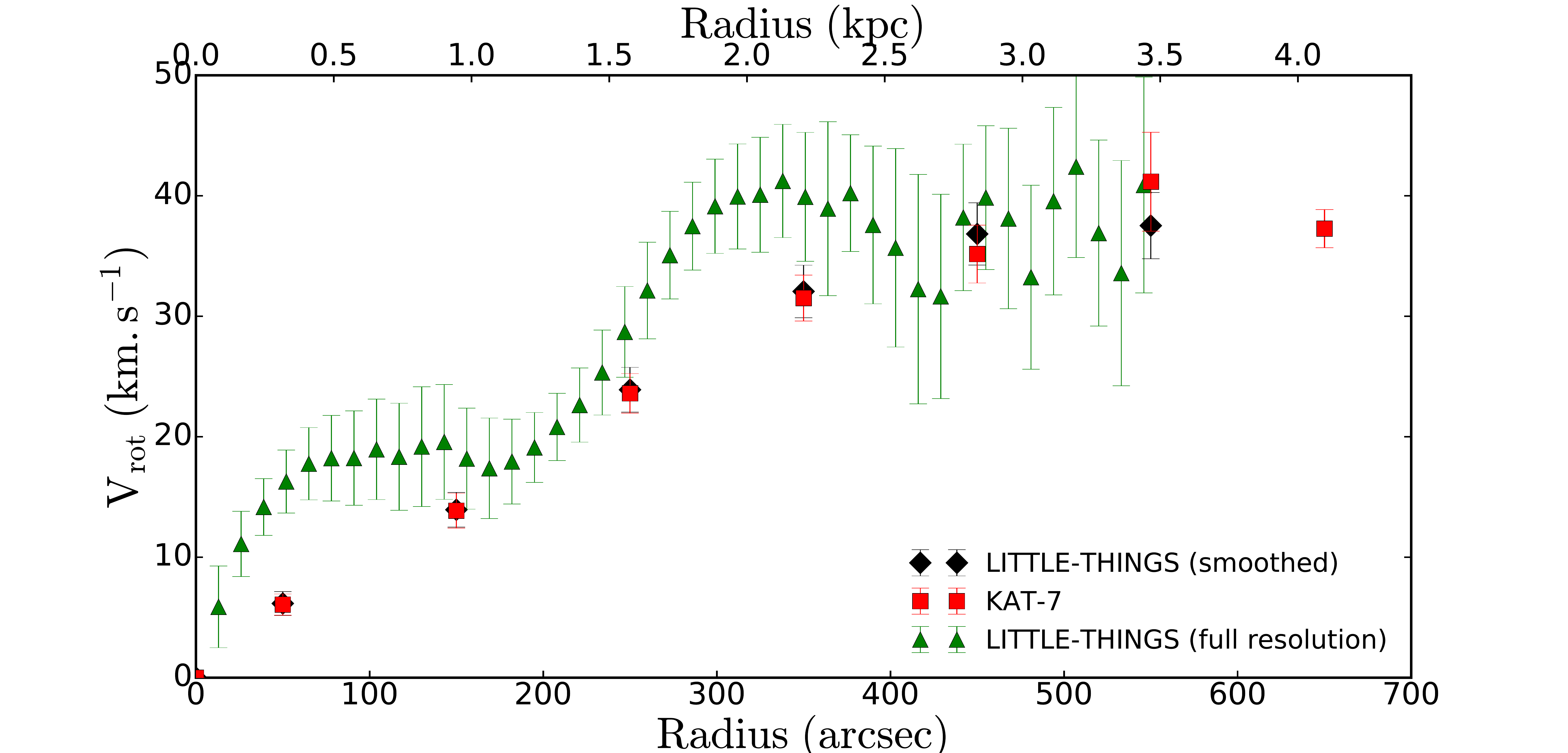}}}
\caption{Comparison of the KAT-7 and VLA LITTLE THINGS rotation curves of Sextans B. The red square show the KAT-7 rotation curve, the black triangles show the LITTLE THINGS RC smoothed to the KAT-7 spatial resolution, and the green triangles show the full resolution LITTLE THINGS RC.}
  \label{fig_speczvsphotz} 
\end{figure}

\begin{table}   
\caption{\small Radial variation of the H\textsc{i} surface densities $\Sigma_{g}$, the gas velocity dispersions $\sigma$, the observed rotation velocities $V_{0}$, the errors of the observed velocities $\Delta V$, and the asymmetric drift corrected rotation velocities $V_{c}$ for the KAT-7 data of Sextans B.}
\begin{minipage}{\textwidth}
\begin{tabular}{l@{\hspace{0.30cm}}c@{\hspace{0.30cm}}c@{\hspace{0.30cm}}c@{\hspace{0.30cm}}c@{\hspace{0.30cm}}c@{\hspace{0.10cm}}}   
\hline

Radius & $\Sigma_{g}$   &  $\sigma $ &  V$_{0}$ & $\Delta$ V  &  V$_{\text{{c}}}$  \\
arcsec & M$_{\odot}$pc$^{-2}$    &   km.s$^{-1}$   &    km.s$^{-1}$ &  km.s$^{-1}$ & km.s$^{-1}$ \\ \hline \hline
0.0&3.7&13.3&0.0&0.0&0.0\\
50&3.5&13.2&6.1&0.9&6.2\\
150&2.7&12.7&14.1&1.5&16.6\\
250&1.8&12.1&24.0&1.5&28.1\\
350&0.9&11.1&32.6&1.0&37.6\\
450&0.4&9.7&37.9&1.3&43.4\\
550&0.2&8.8&38.6&1.8&44.4\\
650&0.1&6.9&37.3&6.0&41.3\\

\hline    
\multicolumn{6}{@{} p{8.5 cm} @{}}{\footnotesize{\textbf{Notes.} Column (1) gives the radius, column (2) the surface densities, column (3) the velocity dispersion, column(4) the observed rotation velocities, column (5) the errors of those velocities, and column (6) the corrected velocities used for the mass models.}}
\label{coords_table}
 
\end{tabular}   

\end{minipage}
\end{table}  

\section{Mass models and dark matter content}
The rotation curve reflects the dynamics of the disk due to the total mass of the galaxy, luminous and dark matter. Dwarf irregulars, like most low-mass surface density galaxies, are believed to be dominated by dark matter at all radii due to the small contribution of luminous matter (stars and gas) to the total dynamics \citep{1989ApJ...347..760C}. The extended H\textsc{i} rotation curves of dwarf irregulars allow us to probe dark matter potentials to much larger radii, making them ideal objects for studying dark matter properties in galaxies. To this end, we decompose the observed rotation curve of Sextans A and B into the luminous and dark matter mass components and verify if dark matter indeed dominates the total dynamics of these systems. The pseudo-isothermal DM halo model (ISO) \citep{1991MNRAS.249..523B} and the Navarro-Frenk-White DM halo model (NFW) \citep{1997ApJ...490..493N} have been used to derive the dark matter components of Sextans A and B. Detailed description of the mass  models has been given in \citet{2017arXiv170809447N}.

\subsubsection{Gas and Stellar components}
We derive the mass models for the stellar and gas components following the procedure described by \citet{2017arXiv170809447N}. The Wide Field Infrared Survey Explorer (WISE) surface brightness profiles in the 3.4 $\mu$m band were used to account for the stellar contribution. The luminosity profiles are from Jarrett (private communication). At 3.4 $\mu$m, WISE probes the emission from the old stellar disk population and is also less affected by dust.

\subsection{Fitting ISO and NFW models for Sextans A and B}
The GIPSY task ROTMAS was used to construct the mass models of Sextans A and B. We fitted the ISO and NFW models to the rotation curves derived from the tilted ring models, taking into account the mass of the luminous matter (stars and gas). All the data points were weighted according to the errors using the inverse square weighting. 

The values of the mass to light ratio, M/L ($\Upsilon*$) were determined by scaling the contribution of the stellar rotation curve to the total rotation curve. This was done using two predetermined M/L values: 1), the M/L of 0.2 \citep{2016AJ....152..157L}. This value has been derived to be the lower end of the mass to light ratio for dwarf galaxies at mid-infrared band, and 
2), the M/L derived using Equation 4 \citep{2014ApJ...782...90C}. This equation is used to calculate the M/L of star forming low mass galaxies. 
\begin{equation}
\log_{10} = M_{\textit{stellar}}/L_{\textit{W1}} = -1.93 (W_{3.4\mu m}- W_{4.6\mu m}) - 0.04
\end{equation}
where W$_{3.4 \mu m}$ - W$_{4.6 \mu m}$ is 0.04 $\pm$ 0.02 for Sextans A and 0.01 $\pm$ 0.02 for Sextans B. This gives the M/L of 0.9 and 0.8 for Sextans A and B respectively.

\subsection{Mass model results for Sextans A}
The fitted parameters for the mass model results of Sextans A are shown in Figure 13 and Table 6. Figure 13 shows that Sextans A is dark matter dominated at all radii. Regardless of the assumption made for the M/L ratio, the ISO model produces a better fit compared to the NFW at radius $\leqslant$ 1.3 kpc. Beyond that radius, the two models tend to give similar fits to the observed rotation curve. For the ISO halo models, we find the lowest $\chi^{2}$ of 3.4 when the M/L value of 0.2 \citep{2016AJ....152..157L} is used. It is not surprising that we get a better fit when we use a smaller M/L of 0.2. Literature shows that for most dwarf galaxies the stellar disk does not contribute significantly to the rotation curve \citep{2003MNRAS.340...12W,2008AJ....136.2648D,2011AJ....141..193O}. The mass models yields DM halo parameters of R$_{0} \sim 0.5$ kpc and $\rho_{0} \sim 0.1$ M$_{\odot}$ pc$^{-3}$. 
\begin{figure}
\centering
\resizebox{1.0\hsize}{!}{\rotatebox{0}{\includegraphics{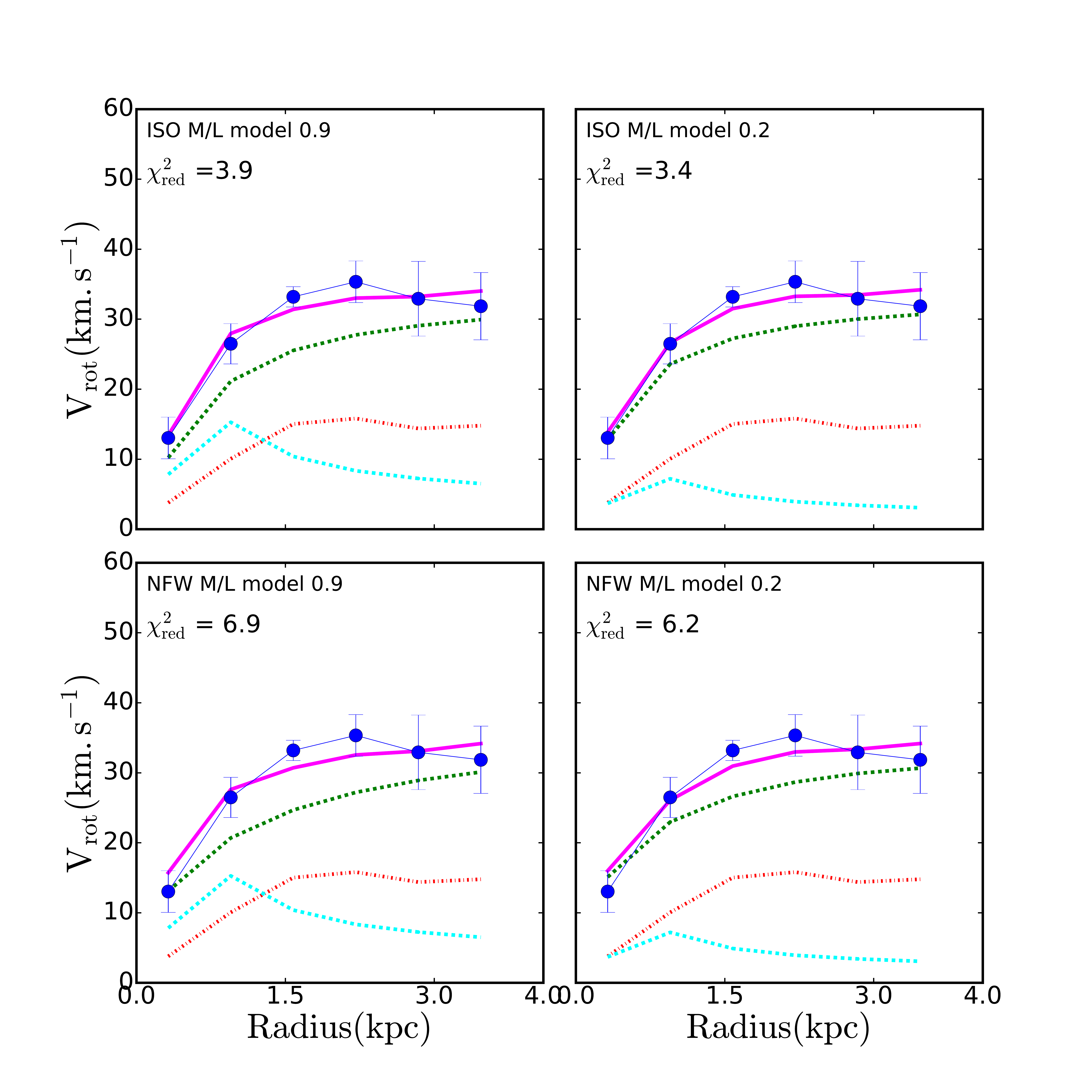}}}
\caption{ISO and NFW mass modeling results of Sextans A. The decomposition of Sextans A rotation curve using two assumption of $\Upsilon_{*}$. The blue circles indicate the observed rotation curve, the magenta lines show the fitted rotation curve, the green dotted lines indicate the dark matter rotation velocities, and the red dot-dashed and blue dashed lines show the rotation velocities of the gas and of the stellar components, respectively.}
  \label{fig_speczvsphotz} 
\end{figure}

\subsection{Mass model results for Sextans B}
The dark matter models for Sextans B are shown in Figure 14 and Table 6. Both ISO and NFW models give larger $\chi^{2}$ values than for the models of Sextans A but again ISO
with an M/L of 0.2 \citep{2016AJ....152..157L} gives the best fit to the data but with a small overestimate for r < 1.3 kpc. This suggests that M/L
 should be slightly smaller. \citet{2015AJ....149..180O} derived a mass model for Sextans B using higher resolution observations. Their results show a better fit with smaller values of $\chi^{2}$ compared to our mass model fits. However, this can be due to the fact that they only fit the rising part of their derived rotation curve, leaving out the outer radii 
were the rotation velocities are declining.

\begin{figure}
\centering
\resizebox{1.0\hsize}{!}{\rotatebox{0}{\includegraphics{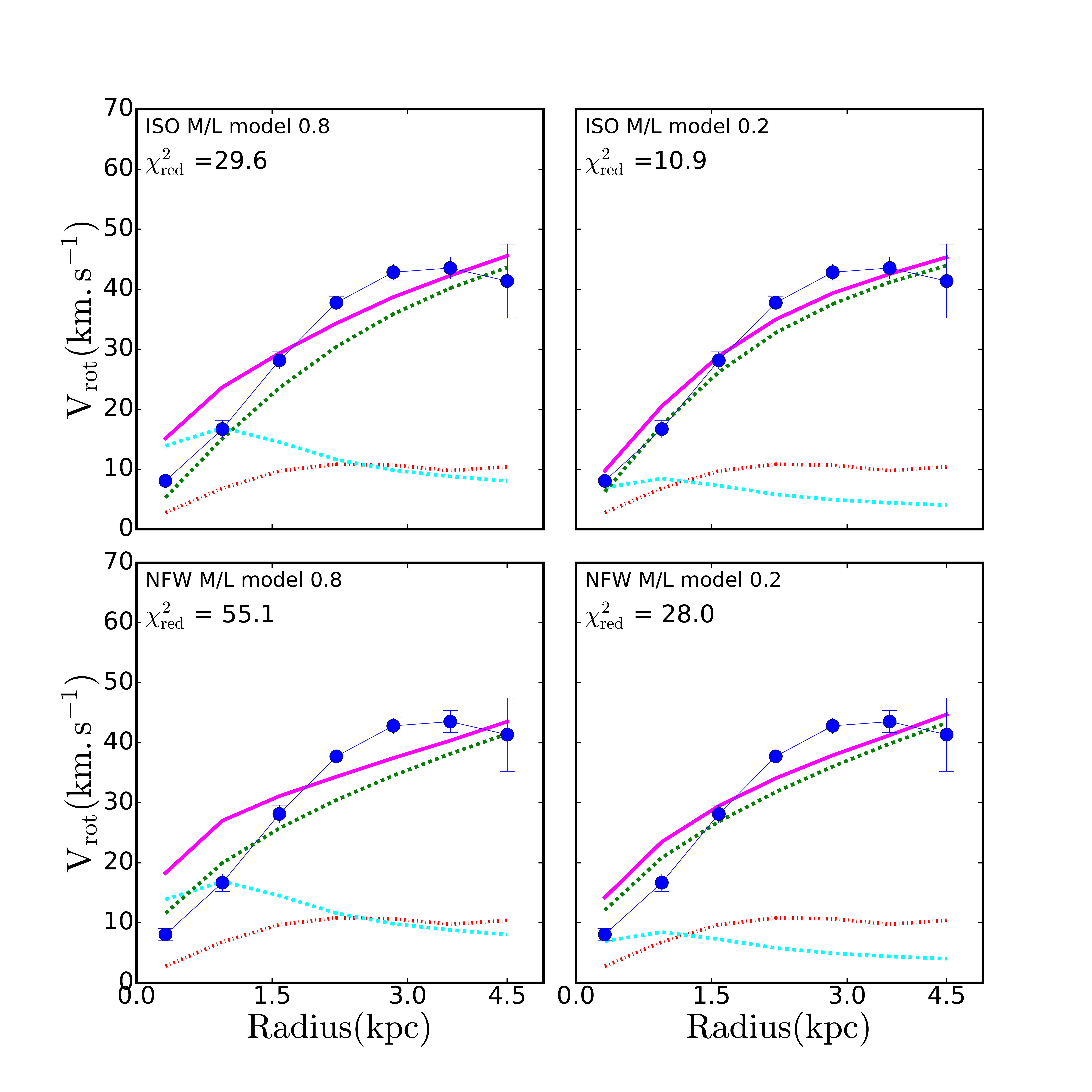}}}
\caption{ISO and NFW mass modeling results of Sextans B. The decomposition of Sextans B rotation curve using two assumption of $\Upsilon_{*}$. The blue circles indicate the observed rotation curve, the magenta lines show the fitted rotation curve, the green dotted lines indicate the dark matter rotation velocities, and the red dot-dashed and blue dashed lines show the rotation velocities of the gas and of the stellar components, respectively.}
  \label{fig_speczvsphotz} 
\end{figure}

\begin{table}
\scriptsize

\caption{\small Results for the Mass Models Sextans A and Sextans B.}
\begin{minipage}{\textwidth}
\begin{tabular}{l@{\hspace{0.6cm}}c@{\hspace{0.6cm}}c@{\hspace{0.5cm}}c@{\hspace{0.5cm}}c@{\hspace{0.5cm}}}   
\hline

   
\hline \hline  
&&Sextans A&& \\
\cline{2-3}
&&ISO Halo&& \\ 
\cline{2-3}
Assumption &$\Upsilon_{*}$ & R$_{0}$ &$\rho_{0}$ &$\chi_{\text{red}}^{2}$\\ 
(1)&(2)&(3)&(4)&(5) \\ \hline
M/L model & 0.90& 0.46 $\pm$ 0.11& 93.57$\pm$ 35.18&3.90\\ 
M/L model & 0.20& 0.40 $\pm$ 0.08& 125.39$\pm$ 44.61&3.40\\ \hline
&&NFW Halo&& \\
\cline{2-3}
Assumption &$\Upsilon_{*}$ & c &R$_{200}$ &$\chi_{\text{red}}^{2}$\\ 
(6)&(7)&(8)&(9)&(10) \\ \hline
M/L model & 0.90& 10.61 $\pm$ 3.49& 35.57$\pm$ 6.98&6.90\\ 
M/L model & 0.20& 12.62$\pm$ 3.71& 33.84$\pm$5.57&6.20\\ \hline

\hline  

&&Sextans B&& \\
\cline{2-3}

&&ISO Halo&& \\
\cline{2-3}
Assumption &$\Upsilon_{*}$ & R$_{0}$ &$\rho_{0}$ &$\chi_{\text{red}}^{2}$\\ 
(11)&(12)&(13)&(14)&(15) \\ \hline
M/L model & 0.80& 2.45 $\pm$ 1.58& 15.21$\pm$ 8.71&29.66\\
M/L model & 0.20& 1.86 $\pm$ 0.57& 21.35$\pm$ 6.96&10.93\\ \hline
&&NFW Halo&& \\
\cline{2-3}
Assumption &$\Upsilon_{*}$ & c &R$_{200}$ &$\chi_{\text{red}}^{2}$\\ 
(16)&(17)&(18)&(19)&(20) \\ \hline
M/L model & 0.80& -0.38 $\pm$ ....& 1494.36$\pm$ ....&55.16\\ 
M/L model & 0.20& -0.35$\pm$ ....& 1518.18$\pm$....&28.04\\ \hline

\multicolumn{5}{@{} p{9.0 cm} @{}}{\footnotesize{\textbf{Notes.} Columns 1, 11 and 6, 16, the stellar $\Upsilon_{*}$ assumption. Column 2, 12 and 7,17, 
$\Upsilon_{*}$ . Column 3, 13, fitted scaling radius of the ISO halo model in kpc. Column 4, 14, fitted central density of the pseudo-isothermal halo model in 10$^{-3}$ M$_{\odot}$ pc$^{-3}$. Column 9, 19 the radius in kpc where the density contrast exceeds 200. Column 8, 18 concentration parameter c of the NFW halo model. Columns 5, 15 and 10, 20, reduced $\chi^{2}$ value. The dotted line (...) are due to unphysical large values of uncertainties.}}
\label{coords_table}
 
\end{tabular}   

\end{minipage}
\end{table}  

\subsection{MOND Models}
The Modified Newtonian Dynamics (MOND) was proposed by \citep {1988ApJ...333..689M} as an alternative to DM. Milgrom postulated that at small accelerations the usual Newtonian dynamics break down and that the law of gravity needs to be modified. MOND claims to be able to explain the mass discrepancies in galaxies with only the contributions of the gas and stellar components without DM. Several authors have been able to explain the mass distributions of galaxies without DM \citep{1991MNRAS.249..523B,1996ApJ...473..117S,2002A&A...393..453B,2014MNRAS.439.2132R}.

The shape of the predicted MOND rotation curve depends on the interpolating function. The standard \citep{1983ApJ...270..384M} and the simple \citep{2005MNRAS.363..603F} interpolating functions are mostly used in the literature. \citet{2005MNRAS.363..603F} claimed that the simple interpolation function gives more realistic M/L values. This is why we have adopted this interpolation function for our analysis.

\subsection{MOND fits for Sextans A and B}
The MOND fitting procedure has two free parameters. We fit the rotation curve with a$_{0}$ fixed to the universal constant of 1.2 $\times 10^{-8}$ cm.s$^{-2}$ \citep{1991MNRAS.249..523B} while letting M/L free for the simple interpolation function. On the other hand, we let the universal constant free while the M/L is fixed to a value as the ones used for the DM models. We note that although the universal constant a$_{0}$ is expected to be the same for all astrophysical objects, observations biases can cause scatter in a$_{0}$ which can lead to a significant departure of the universal constant from the standard value.

The MOND results for Sextans A and B are given in Figure 15 and 16. Table 7 shows the derived MOND parameters of the two galaxies. The discrepancies between the observed rotation curve and the RC predicted by MOND are visible in both Sextans A and B. The fits give high values of the reduced $\chi^{2}$, with better fits derived when a$_{0}$ is allowed to vary. This poses a challenge as a$_{0}$ should be a universal constant. The best result is obtained for Sextans A with a$_{0}$ free. However, this model suggests a value of a$_{0}$ much smaller than the standard value. 

\begin{figure}
  \includegraphics[width = \columnwidth]{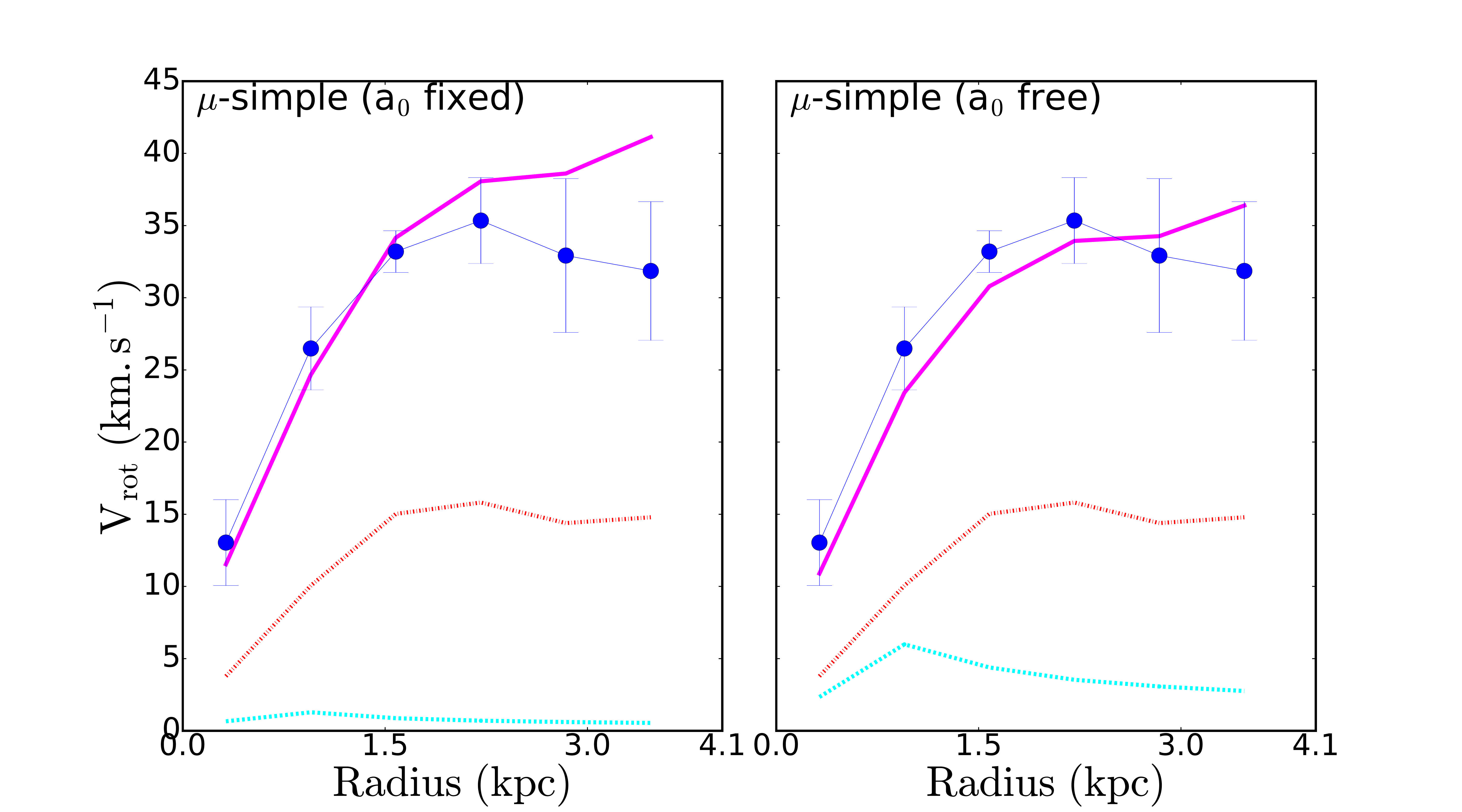}
  \caption{MOND mass models for Sextans A with a$_{0}$ fixed (left) and a$_{0}$ free (right) for the simple interpolation function. The red dashed curve is for the H\textsc{i} disk, the dash-dotted light blue curve is for the stellar disk, and the continuous purple curve is the MOND contribution. } 
  \label{fig_speczvsphotz} 
\end{figure}

\begin{figure}
  \includegraphics[width = \columnwidth]{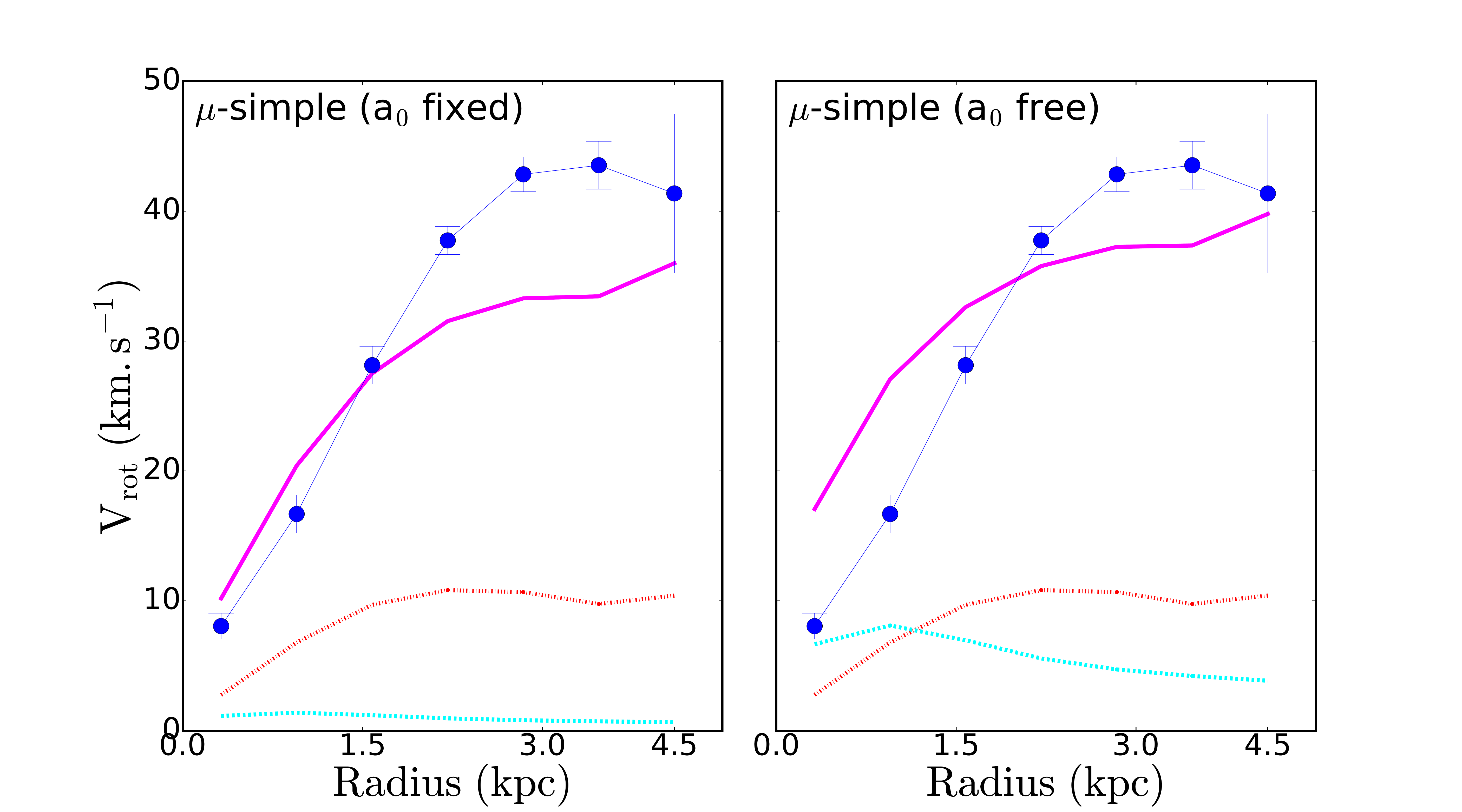}
  \caption{MOND mass models for Sextans B with a$_{0}$ fixed (left) and a$_{0}$ free (right) for the simple interpolation function. The red dashed curve is for the H\textsc{i} disk, the dash-dotted light blue curve is for the stellar disk, and the continuous purple curve is the MOND contribution.} 
  \label{fig_speczvsphotz} 
\end{figure}

\begin{table}   

\caption{\small Results for the MOND Models of Sextans A and B for the KAT-7 Data.}
\begin{minipage}{\textwidth}
\begin{tabular}{l@{\hspace{0.90cm}}c@{\hspace{0.90cm}}c@{\hspace{0.90cm}}c@{\hspace{0.90cm}}}   
\hline 
&Sextans A& \\
\cline{2-2}

a$_{0}$ (cm.s$^{-2}$) &  Parameter& Result \\
& & &\\ \hline \hline
Fixed&(M/L)&0.006\\ 
(1.2 $\times 10^{-8}$) & &\\
& $\chi^{2}_{\text{red}}$& 26.40\\ 
\multicolumn{2}{c}{}&\\
\cline{1-4}
Free& (M/L)& 0.2 \\
(6.9 $\times 10^{-9}$)& & \\
&$\chi^{2}_{\text{red}}$&6.40\\
\multicolumn{2}{c}{}&\\ \hline

&Sextans B& \\
\cline{2-2}

a$_{0}$ (cm.s$^{-2}$) &  Parameter& Result \\
& & &\\ \hline \hline
Fixed&(M/L)&0.005\\
(1.2 $\times 10^{-8}$) & &\\ 
& $\chi^{2}_{\text{red}}$& 46.50\\ 
\multicolumn{2}{c}{}&\\
\cline{1-4}
Free& (M/L)& 0.2 \\
(1.5 $\times 10^{-8}$) & &\\
&$\chi^{2}_{\text{red}}$&47.44\\
\multicolumn{2}{c}{}&\\ \hline
\label{coords_table}
 \end{tabular}   
\end{minipage}
\end{table}  

\subsection{Comparison with the literature}
There are a lot of studies on the dark matter distribution in dwarf galaxies in the literature, but this Section will focus on the comparison between this work and that of \citet{2011AJ....141..193O}, which looks at the dark matter distribution of the LITTLE THINGS galaxies. A dominant trend in the core-like distribution has been reported by \citet{2011AJ....141..193O}. This is in agreement with our mass model results in which the ISO model produces better fits as compared to the NFW model. The KAT-7 ISO model fits suggest a dark matter core radius of R$_{c}$ = 0.46 to 2.45 kpc and density $\rho_{0}$ = 15 to 125 M$_{\odot}$ pc$^{-3}$. The values are in the range of \citet{2011AJ....141..193O}. \citet{2011AJ....141..193O} found the values of R$_{c}$ and $\rho_{0}$ to be in the range 0.10 to 9.82 kpc and 1.8 to 725 M$_{\odot}$ pc$^{-3}$ respectively.
 
 For both observations, we see cases were the NFW fit yields unrealistic halo parameters when trying to fit the observed RCs. From our KAT-7 data, we obtained c values < 0 for Sextans B while the LITTLE THINGS observations record c values < 0.1 for DD0 53 and NGC 2366. As emphasized by \citet{2008AJ....136.2648D}, c represents the collapse factor, and therefore c values < 1 make no sense in the CDM context. The estimated concentration parameter for Sextans A is $\sim$ 11. The LITTLE THINGS observations record c values ranging from $\sim$ 4 to 48. The concentration parameter for Sextans A is within the LITTLE THINGS range and also consistent with c values derived for most dwarf galaxies \citep{2008AJ....136.2648D}. We obtain V$_{200}$ of 26 km.s$^{-1}$
for Sextans A. This value is in the range of  \citet{2011AJ....141..193O}. \citet{2011AJ....141..193O} record V$_{200}$ values ranging from 3.7 to 58 km.s$^{-1}$.

One thing to note from our KAT-7 mass model fits is that although we do not get perfect fits to our observed RCs, the parameters we derived for Sextans A and B are in agreement and within the ranges found for other dwarf galaxies.

\section{Star formation thresholds}
The mechanisms that assist star formation activities in low surface brightness (LSB) galaxies remain a puzzling question. The low stellar densities in most gas rich LSB galaxies imply that the star formation process has been inefficient in converting the available gas into stars. Several studies \citep{1996ApJ...462..732H,1998ApJ...493..595H,2006AJ....131..363D} have been able to show that, although being gas rich, the gas densities in most LSB galaxies fall below the threshold needed to support star formation. 

\citet{1989ApJ...344..685K} found that a modified Toomre-Q criterion model, could satisfactorily describe the star formation threshold gas densities in active star forming galaxies. In this model, the density above which the gas becomes unstable and form stars is a function of the kinematics of the galaxy. The threshold for star formation depends on the parameter $\alpha_{c}$, which is defined by \citep{2001ApJ...555..301M}:
\begin{equation}
\alpha_{c} = \frac{\Sigma_{c}}{\Sigma_{H\textsc{i}}}
\end{equation}
where $\Sigma_{HI}$ is the gas surface density corrected for helium and other metals, and $\Sigma_{c}$ is the critical density for cloud formation. The critical density is described by \citep{2001ApJ...555..301M}: 
\begin{equation}
\Sigma_{c}(r) = \alpha \frac{\sigma(r) k(r)}{\pi G}
\end{equation}
where $\sigma_{r}$ is the gas velocity dispersion, $\alpha$ is a constant close to unity which is included to account for a more realistic
disk, and $k(r)$ is the epicyclic frequency given by 
\begin{equation}
k^{2} (r) = 2\big(\frac{V^{2}}{r^{2}} + \frac{V}{r} \frac{dV}{dr})
\end{equation}
where V is the rotation velocity in kpc, and r is the radius in km. \citet{1989ApJ...344..685K} found $\alpha_{c}$ = 0.63 at the edge of star forming disks and \citet{2001ApJ...555..301M} found $\alpha_{c}$ = 0.69. If $\Sigma_{g}(r)$, the surface density of the gas in the disk exceed $\Sigma_{c}(r)$, then the disk will be unstable to axisymmetric disturbances and large scale star formation can occur. 

An alternative to the Toomre-Q criteria is the cloud growth criteria based on shear. This criteria explains the star formation threshold based on the local shear rate \citep{1998ApJ...493..595H} and is described by the Oort's A constant:

\begin{equation}
A = -0.5 \times r\frac{d\Omega}{dr} = 0.5\big(\frac{V}{r} - \frac{dV}{dr})
\end{equation}
Then the threshold has the form
\begin{equation}
\Sigma_{c,A}(r)= \frac{\alpha(r) \sigma(r) A}{\pi G}
\end{equation}
Where $\alpha_{c,A}$ is taken to be 2.5 \citep{1998ApJ...493..595H}, but the normalization for $\Sigma_{c,A}$ is relatively uncertain. In this case, the threshold parameter $\alpha$ is defined by
\begin{equation}
\alpha_{c,A} = \frac{\Sigma_{H\textsc{i}}}{\Sigma_{c,A}}
\end{equation}

We have used the KAT-7 observations to examine the star formation threshold throughout Sextans A and B as a function of radius to determine if the subcritical gas density is preventing the galaxies from large scale star formation. To derive the critical densities, a constant gas velocity dispersion $\sigma$ of 9 km.s$^{-1}$ is used. This is derived from the medium value of our azimuthally averaged H\textsc{i} velocity dispersion radial profile. We have assumed $\alpha$ = 1 and establish the ratios of the gas densities to the critical densities as a function of radii using the Toomre-Q and cloud-growth based on shear criteria. 

\subsection{Star formation threshold results for Sextans A and B}
Figure 17 (a) and (b)  compares the observed H\textsc{i} gas surface densities to the critical densities derived using the Toomre-Q \citep{1989ApJ...344..685K} and the cloud-growth based on shear \citep{1998ApJ...493..595H} criterion for Sextans A and B respectively. The gas surface densities in Sextans A and B are low relative to the Toomre-Q critical densities $\Sigma_{c}$ necessary for instabilities that lead to cloud formation and later star formation. We find that the critical densities $\Sigma_{c,A}$ in Sextans A exceed the gas surface density at all radii. For Sextans B, $\Sigma_{c,A}$ is lower than the gas surface density in regions $\leqslant$ 2.1 kpc. 

Figure 18 and 19 shows the radial dependance of the ratios $\Sigma_{H\textsc{i}}/\Sigma_{c}$ and $\Sigma_{H\textsc{i}}/\Sigma_{c,A}$ for Sextans A and B. We see that for both galaxies, the ratios derived from the Toomre-Q critical densities $\Sigma_{H\textsc{i}}/\Sigma_{c}$ fail to predict the observed star formation in Sextans A and B. At all radii, $\Sigma_{H\textsc{i}}/\Sigma_{c}$ is below the stability parameter $\alpha$ = 0.63, the median value from \citet{1989ApJ...344..685K} above which the gas density is high enough for large scale star formation.  We calculate a mean $\alpha_{c}$ = 0.25 and 0.20 for Sextans A and B by finding the ratio $\Sigma_{g}/\Sigma_{c}$ from the center to the Holmberg radius. This result suggests that the gas surface density in Sextans A and B is low to effectively form stars if the Toomre-Q stability criterion is used to determine the star formation in these galaxies.  

The ratio $\Sigma_{H\textsc{i}}/\Sigma_{c,A}$ seems better suited to explain the star formation in Sextans A and B. In both galaxies, $\Sigma_{H\textsc{i}}/\Sigma_{c,A}$ exceeds the stability parameter $\alpha$ = 0.63 in the inner regions. We measure a mean $\alpha_{c,A}$ value of 0.64 and 5 for Sextans A and B respectively. These results suggest that shear may play an important role in facilitating cloud formation and later induce star formation in Sextans A and B. 

Many different models have been used to examine the star formation threshold in dwarf galaxies. \citet{1998ApJ...493..595H} used different models to determine regions of star formation using a sample of spiral and dwarf galaxies. Their results showed that for dwarf galaxies, the Toomre-Q ratio $\alpha_{c}$ was too low to induce star formation at all radii. Examples of their derived average $\alpha_{c}$ values include: 0.45 for IC 1613, 0.36 for DD0 154, 0.30 for DD0 155, and 0.26 for Sextans A. These values are in excellent agreement with our calculated $\alpha_{c}$ values of Sextans A and B using the KAT-7 observations. We reach the same conclusion as \citet{1998ApJ...493..595H} that the cloud-growth criterion based on shear is well suited to explain star formation in star forming regions of dwarf galaxies. In most cases, \citet{1998ApJ...493..595H} derived $\alpha_{c,A}$ values close to 1. We see that the high values of $\Sigma_{H\textsc{i}}/\Sigma_{c,A}$ derived for Sextans B are not odd when compared with the literature. \citet{1998ApJ...493..595H} records high $\Sigma_{H\textsc{i}}/\Sigma_{c,A}$ for IC 1613 and DD0 50, showing the highest peak $\sim$ 10. 

\begin{figure}
\centering
   \subcaptionbox{Sextans A\label{fig3:b}}{\includegraphics[width=3.0in]{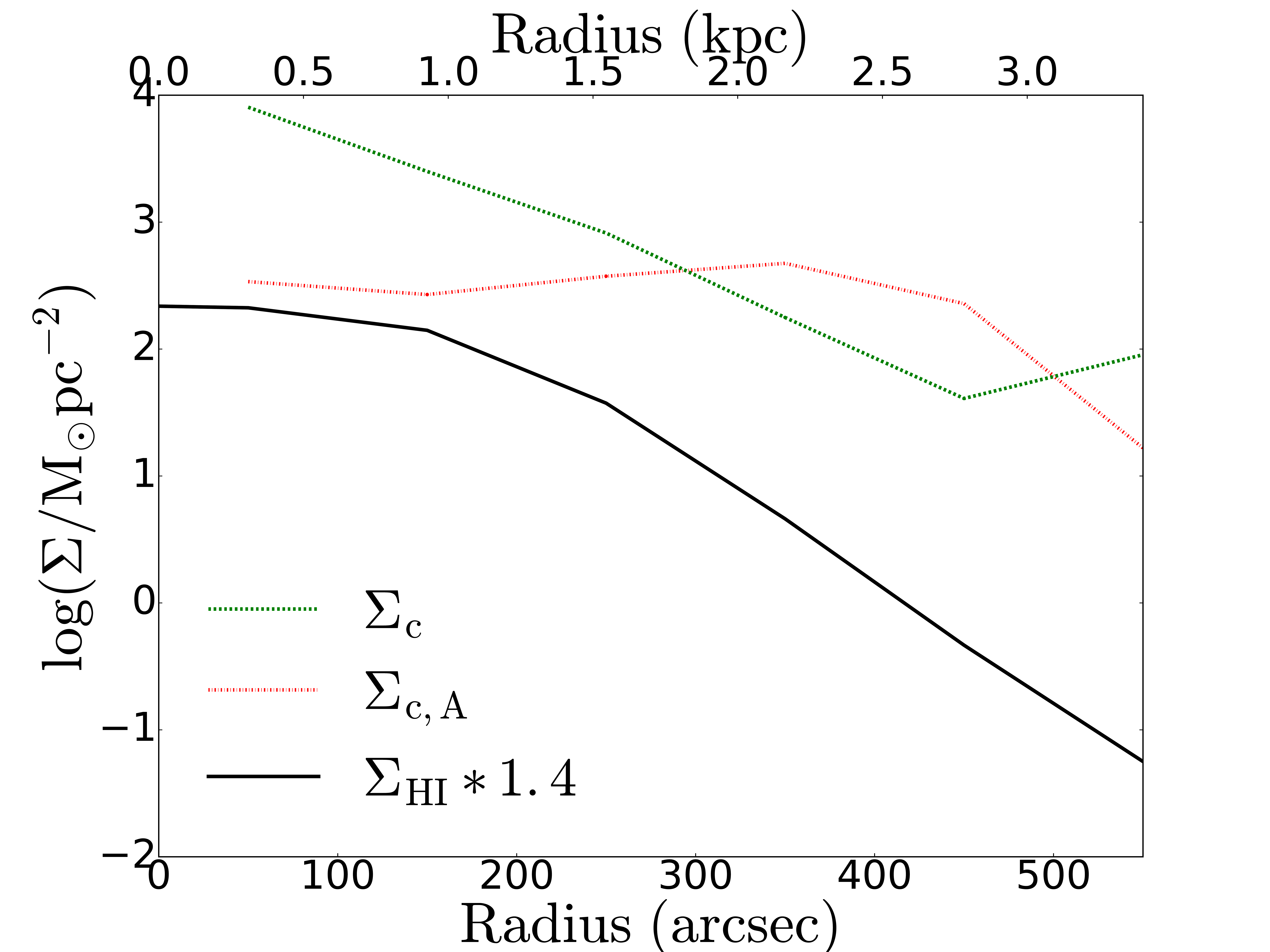}} \hspace{-1mm}  
   \subcaptionbox{Sextans B\label{fig3:a}}{\includegraphics[width=3.0in]{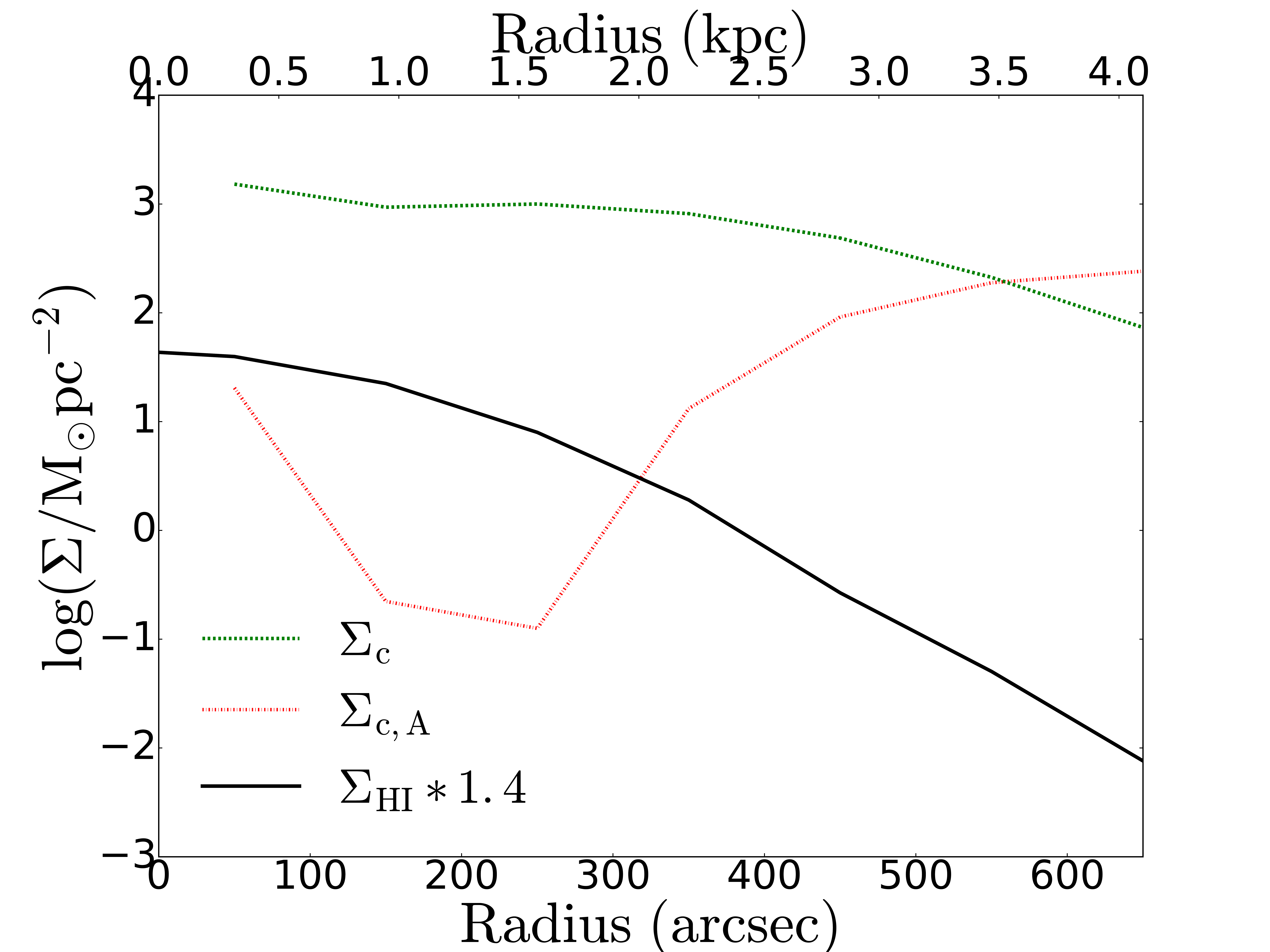}}\hspace{-1mm}\\%
  
   \caption{Shows the gas surface density, black solid line (corrected for helium), the green dashed line shows the critical density derived using the \citet{1989ApJ...344..685K}, and the red dash-dotted line shows the critical density as derived by \citet{1998ApJ...493..595H}.}  
   \end{figure}

\begin{figure}
\centering
   \includegraphics[width=3.0in]{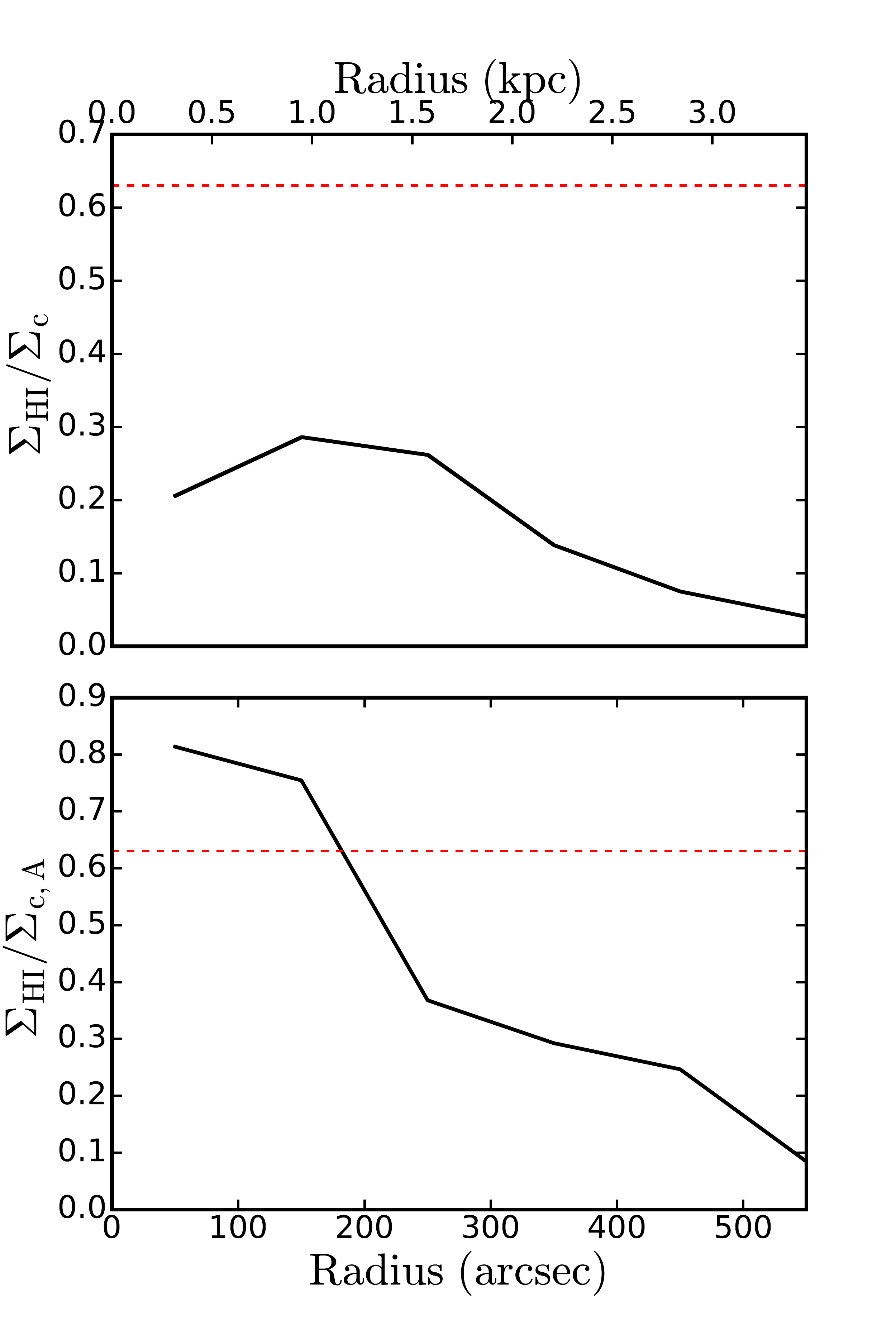} \hspace{5mm}\\%
  
   \caption{The radial variation in the ratios of the observed gas surface density to the critical densities for Sextans A. The top panel shows the ratio $\alpha_{c}$ = $\Sigma_{H\textsc{i}/\Sigma_{c}}$ calculated using the Toomre-Q while the bottom panel shows the ratio $\alpha_{c,A}$ = $\Sigma_{H\textsc{i}/\Sigma_{c,A}}$ calculated using cloud-growth criteria based on shear. The red dashed line shows $\alpha_{Q}$ = 0.63, the median value from \citet{1989ApJ...344..685K} above which the gas density is high enough for large scale star formation.}  
   \end{figure}

 \begin{figure}
\centering  
   \includegraphics[width=3.0in]{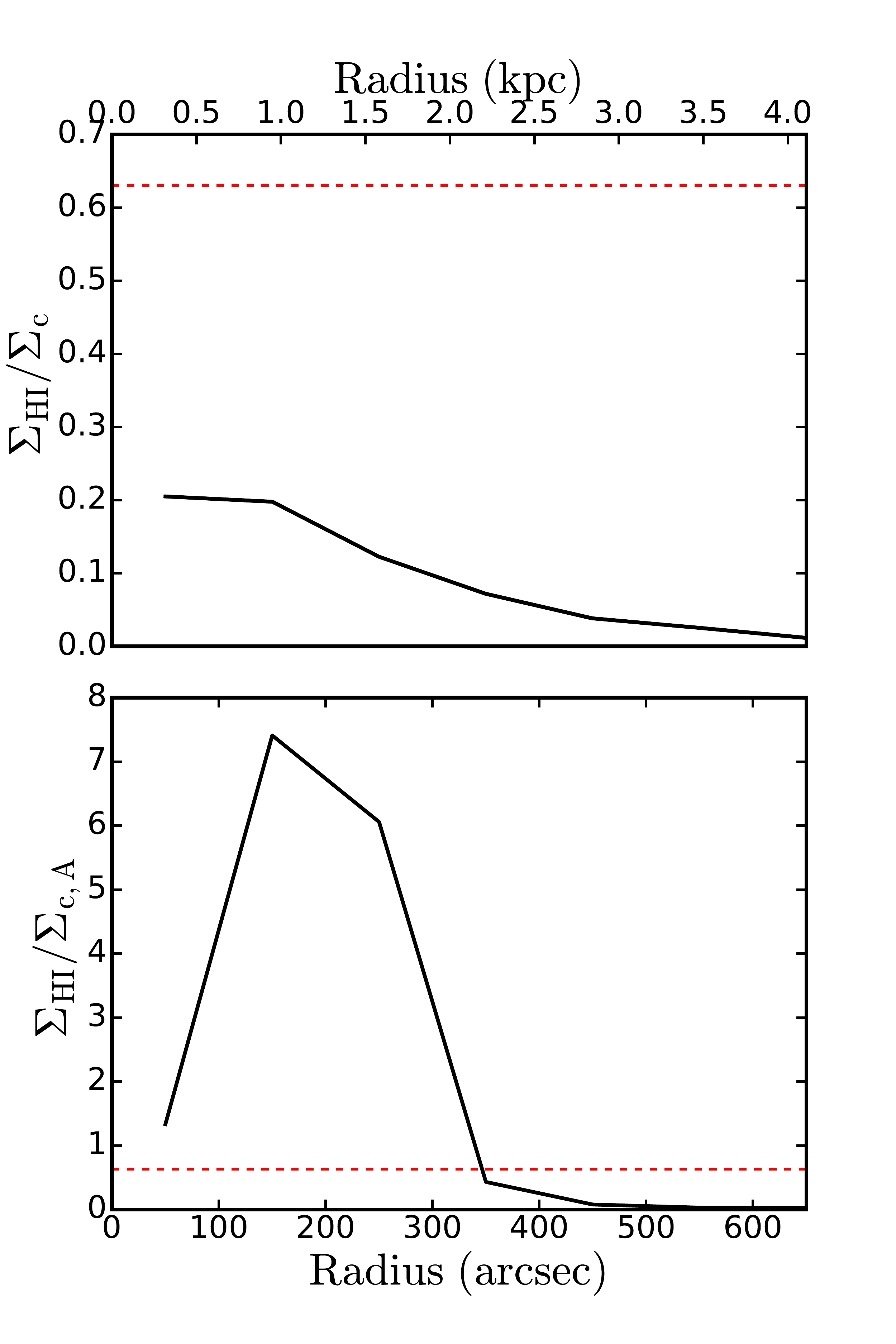}\hspace{5mm}\\%
  
   \caption{The radial variation in the ratios of the observed gas surface density to the critical densities for Sextans B. The top panel shows the ratio $\alpha_{c}$ = $\Sigma_{H\textsc{i}/\Sigma_{c}}$ calculated using the Toomre-Q while the bottom panel shows the ratio $\alpha_{c,A}$ = $\Sigma_{H\textsc{i}/\Sigma_{c,A}}$ calculated using cloud-growth criteria based on shear. The red dashed line shows $\alpha_{Q}$ = 0.63, the median value from \citet{1989ApJ...344..685K} above which the gas density is high enough for large scale star formation.}  
   \end{figure}

\section{Summary and conclusion}

We have obtained high sensitivity, intermediate resolution H\textsc{i} observations of Sextans A and B and have used them to study the H\textsc{i} distribution, kinematics, and star formation thresholds. At column densities of 5.8 and 5.4 $\times 10^{18}$ cm$^{-2}$ for Sextans A and B, we do not detect new H\textsc{i} emission. In fact, our results, which are close to the GBT and VLA, contradicts the large extents and fluxes claimed by \citet{1981A&A...102..134H} from the Effelsberg observations. 

A tilted ring model is fitted to the H\textsc{i} velocity fields to derive the rotation velocities. We measure the RCs of Sextans A and B out to 3.5 and 4.0 kpc respectively. The RCs of both galaxies are seen to decline at outer radii. For Sextans A, we calculate a mean V$_{sys}$ = 342 $\pm$ 0.6 km.s$^{-1}$, P.A. = 34$^{\circ}$, and inclination = 86$^{\circ}$ while V$_{sys}$ = 302 $\pm$ 0.9 km.s$^{-1}$, P.A. = 57$^{\circ}$, and inclination = 49$^{\circ}$ are calculated for Sextans B.

Using the observed RCs as mass model inputs show that the galaxies have a higher fraction of dark matter compared to luminous matter (gas and stars).
For Sextans A, the ISO DM model reproduces better the observed RC than the NFW model. In the case of Sextans B both DM models fail to represent properly the data but with
a slightly better fit for the ISO model with a M/L of 0.2. Better mass model fittings are obtained when the stellar disk does not contribute significantly to the rotation curve. Fixing the M/L to a predetermined value of 0.2 \citep{2016AJ....152..157L} produces better fits than using high M/L values of 0.9 and 0.8 for Sextans A and B derived in the infrared band using the WISE colors. For both galaxies, the MOND model produces better fit when a$_{0}$ is allowed to vary. This poses a challenge as a$_{0}$ should be considered as a universal constant. 

Regions of star formation in Sextans A and B are better explained using the cloud-growth criterion based on shear as compared to the Toomre-Q criterion, as was found previously for NGC 6822
\citep{2017arXiv170809447N}

\section*{ACKNOWLEDGEMENT}
We thank the entire SKA SA team for allowing us to obtain scientific data during the commissioning phase of KAT-7. We thank Tom Jarrett for providing the WISE luminosity profiles.
 This research is supported by the South 
African Research Chairs Initiative (SARCHI) of the Department of Science and Technology (DST), the Square Kilometer Array South Africa (SKA SA) and the 
National Research Foundation (NRF).




\bibliographystyle{mn2e}





\bsp	
\label{lastpage}
\end{document}